\documentclass{aa}
\usepackage{graphicx}
\usepackage{txfonts}

\begin{document}

\title{Two-dimensional spectroscopy of a sunspot}
\subtitle{III. Thermal and kinematic structure of the penumbra
at 0\farcs5 resolution}

\author{L.R.\ Bellot Rubio\inst{1,2}, R.\ Schlichenmaier\inst{2}, 
\and A.\ Tritschler\inst{3,2} }
\institute{Instituto de Astrof\'{\i}sica de Andaluc\'{\i}a (CSIC), 
Apdo.\ 3004, 18008 Granada, Spain \\
\email{lbellot@iaa.es}
\and
Kiepenheuer-Institut f\"ur Sonnenphysik, 
Sch\"oneckstr. 6, 79104, Freiburg, Germany 
 \and
National Solar Observatory/Sacramento Peak\thanks{Operated by
the Association of Universities for Research in Astronomy, Inc.\ (AURA), 
for the National Science Foundation}, P.O.\ Box 62, Sunspot, 
NM 88349, USA}

\date{Received / Accepted }

\abstract{We investigate the thermal and kinematic configuration of a sunspot
penumbra using very high spectral and spatial resolution intensity profiles of
the non-magnetic \ion{Fe}{i} 557.6 nm line. The dataset was acquired with the
2D solar spectrometer TESOS. The profiles are inverted using a one-component
model atmosphere with gradients of the physical quantities. From this
inversion we obtain the stratification with depth of temperature,
line-of-sight velocity, and microturbulence across the penumbra. Our results
suggest that the physical mechanism(s) responsible for the penumbral filaments
operate preferentially in the lower photosphere. We confirm the existence of a
thermal asymmetry between the center and limb-side penumbra, the former being
hotter by 100-150 K on average. We also investigate the nature of the bright
ring that appears in the inner penumbra when sunspots are observed in the
wing of spectral lines. It is suggested that the bright ring does not reflect a 
temperature enhancement in the mid photospheric layers. The line-of-sight
velocities retrieved from the inversion are used to determine the flow speed
and flow angle at different heights in the photosphere. Both the flow speed
and flow angle increase with optical depth and radial distance.  Downflows are
detected in the mid and outer penumbra, but only in deep layers ($\log
\tau_{500} \leq -1.4$). We demonstrate that the velocity stratifications
retrieved from the inversion are consistent with the idea of penumbral flux
tubes channeling the Evershed flow. Finally, we show that larger Evershed
flows are associated with brighter continuum intensities in the inner
center-side penumbra. Dark structures, however, are also associated with
significant Evershed flows. This leads us to suggest that the bright and dark
filaments seen at 0\farcs5 resolution are not individual flow channels, but a
collection of them. Our analysis highlights the importance of very high
spatial resolution spectroscopic and spectropolarimetric measurements for a better
understanding of sunspot penumbrae.  \keywords{Line: profiles -- Sun:
photosphere -- Sun: sunspots }}

\titlerunning{Thermal and kinematic structure of the penumbra}
\authorrunning{Bellot Rubio et al.}
\maketitle

\section{Introduction}
The fine structure of the penumbra is intimately connected to the
Evershed flow, the most conspicuous dynamical phenomenon observed in
sunspots (e.g., Solanki 2003; Thomas \& Weiss 2004). Therefore, a
proper observational characterization of this fine structure is
essential to understand the Evershed flow and, more generally, the
nature of the penumbra itself.

Studying the fine structure of the penumbra is difficult because of
the very small scales involved. Recent imaging observations with the
Swedish 1-m Solar Telescope (Scharmer et al.\ 2003; Rouppe van der
Voort et al.\ 2004) have revealed that even at a resolution of
0\farcs1 some of the filaments that form the penumbra may be
unresolved. That is, different structures (each having different
properties) may coexist in the resolution element, and we are not able
to separate them. A proper characterization of the penumbra calls not
only for diffraction-limited imaging observations, but also for more
quantitative spectroscopic and spectropolarimetric analyses at the
highest resolution possible.  By interpreting the shape of spectral
lines we can derive the thermal, magnetic, and kinematic
configuration of the penumbra via the Zeeman and Doppler effects.
Unfortunately, current spectroscopic observations do not attain
resolutions better than $\sim 0\farcs5$, and this limit is
often reached at the cost of poor spectral resolution. The situation
is even worse in the case of spectropolarimetric measurements: the
best angular resolution of existing solar polarimeters is about 1\arcsec.

This paper is the third of a series devoted to the analysis of 2D
spectroscopic observations of a sunspot. The data were acquired with the
TElecentric SOlar Spectrometer (TESOS; Kentischer et al.\ 1998; Tritschler et
al.\ 2002) and the Kiepenheuer Adaptive Optics System (KAOS; Soltau et al.\
2002; von der L\"uhe et al.\ 2003) at the German Vacuum Tower Telescope of
Observatorio del Teide (Tenerife, Spain). These observations combine high
spectral ($\lambda/\Delta \lambda \sim 250\,000$) and spatial (0\farcs5)
resolution (Tritschler et al.\ 2004; hereafter referred to as Paper I). In
Paper II of this series (Schlichenmaier et al.\ 2004), we investigated the
properties of the Evershed flow through a bisector analysis of the observed
intensity profiles.

Here we perform a full inversion of the same data set in order to
study the thermal and kinematic configuration of the penumbra. Our
analysis allows us to explain a number of findings reported in Paper~I
and to remove some of the uncertainties associated with the methods
employed in Paper~II.  At the same time, we gather new information on
the thermal properties of the penumbra. This is particularly important
because very few thermal studies of sunspot penumbrae have been
published to date (del Toro Iniesta et al.\ 1994; Balasubramaniam 2002; 
Rouppe van der Voort 2002; S\'anchez Cuberes et al.\ 2005).

A brief account of the observations and details of the inversion
procedure are given in Sect.~\ref{obs}. In Sect.~\ref{results} we
discuss the maps of physical quantities inferred from the
inversion. The thermal and kinematic configuration of the penumbra is
described in detail in Sects.~\ref{thermal} and \ref{kinematic}. Some
implications of the results are presented in Sect.~\ref{discusion}, and
a summary of our findings is given in Sect.~\ref{summary}.

\section{Observations and data analysis}
\label{obs}
\subsection{Observations}
On July 5, 2002, TESOS was used to measure the intensity profiles of
\ion{Fe}{i}~557.6~nm in the main sunspot of NOAA AR~10019 (cf.\ Paper
I). At the time of the observations, the spot was located 23$^{\rm o}$
off the disk center. The line was scanned at 100 wavelength points
with a wavelength step of 8.4 pm, corresponding to overcritical
spectral sampling. The time needed to complete the scan was 37
s. Operated in high resolution mode, TESOS yielded a pixel size of
$0\farcs089 \times 0\farcs089$. During the scan, the KAOS system 
provided real-time correction of wavefront distortions due
to turbulence in the Earth's atmosphere.

\subsection{Diagnostic capabilities of \ion{Fe}{i} 557.6 nm}
We use the intensity profiles of \ion{Fe}{i}~557.6~nm to derive the
stratification of temperature and velocity in the penumbra. Therefore,
it is important to examine the properties of this line in some
detail. \ion{Fe}{i}~557.6 nm is a fairly strong photospheric line with
zero effec\-ti\-ve Land\'e factor, i.e., it does not undergo any
Zeeman splitting. The line is often quoted to be temperature
insensitive, but this is too a simplistic statement. Figure~5 of
Paper~I shows that the equivalent width of \ion{Fe}{i}~557.6 nm
changes by non-negligible amounts when going from the quiet sun to the
umbra, where the temperatures are rather different.

\begin{figure}
\begin{center}
\resizebox{.96\hsize}{!}{\includegraphics{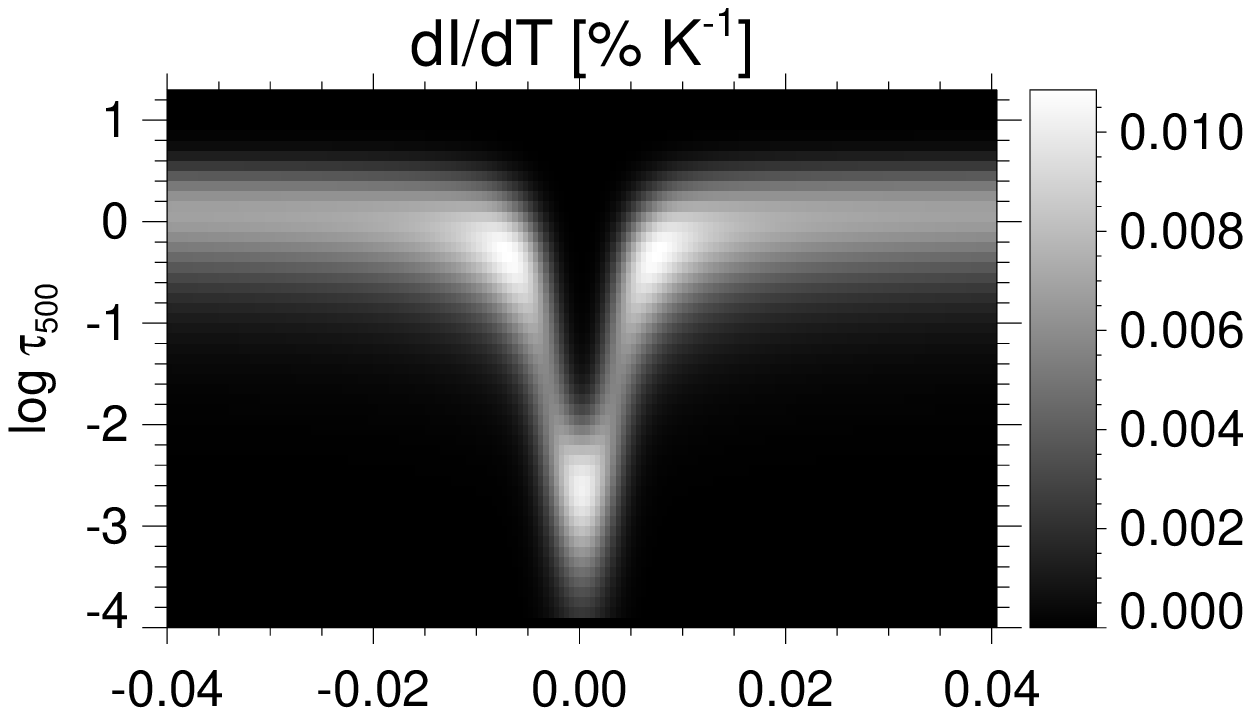}}
\resizebox{.96\hsize}{!}{\includegraphics{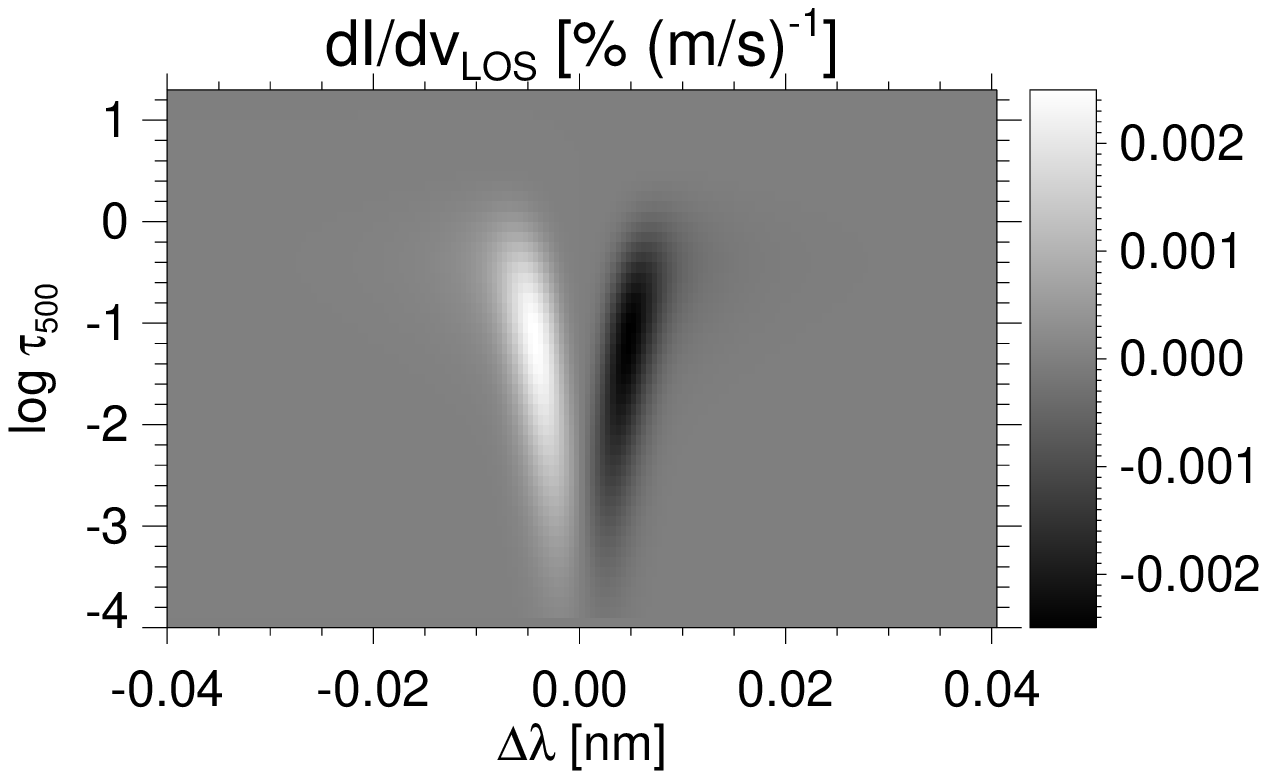}}
\end{center}
\caption{Response functions of the \ion{Fe}{i} 557.6 nm line to
temperature {\em (top)} and line-of-sight velocity {\em (bottom)}. The two plots
indicate the percentage by which the specific intensity at $\lambda$
changes when the temperature (resp.\ LOS velocity) is increased by 1~K 
(resp.\ 1~m~s$^{-1}$) in a layer of optical width $\Delta \log
\tau_{500} = 0.1$ and optical depth $\tau_{500}$.  The thermal
stratification of the penumbral model of TTR has been employed for
these calculations.
\label{rfs} }
\end{figure}

In the following, we elaborate on the temperature sensitivity of
\ion{Fe}{i}~557.6 nm by means of response functions (RFs; see Ruiz Cobo \& del
Toro Iniesta 1994 and del Toro Iniesta 2003).  The upper panel of
Fig.~\ref{rfs} shows the RF of \ion{Fe}{i}~557.6~nm to temperature
perturbations as evaluated in the penumbral model atmosphere of del Toro
Iniesta et al.\ (1994; hereafter TTR). This plot tells us how much the
specific intensity at $\lambda$ changes when the temperature of the model is
increased by 1\,K at a given optical depth $\tau_{500}$. The
figure demonstrates that the line reacts to temperature changes. The
sensitivity of any spectral line to temperature perturbations is determined by
two competing effects: changes in the absorption and in the emission. Usually,
only opacity variations are considered (e.g., Gray 1988, Chapter 13) but, as
pointed out by Cabrera Solana et al.\ (2005), the {\em dominant} contribution
is that of the source function.  Neglect of this dominant factor is the likely
cause of the line being quoted as temperature insensitive in most
publications.

\begin{figure}
\begin{center}
\resizebox{.95\hsize}{!}{\includegraphics{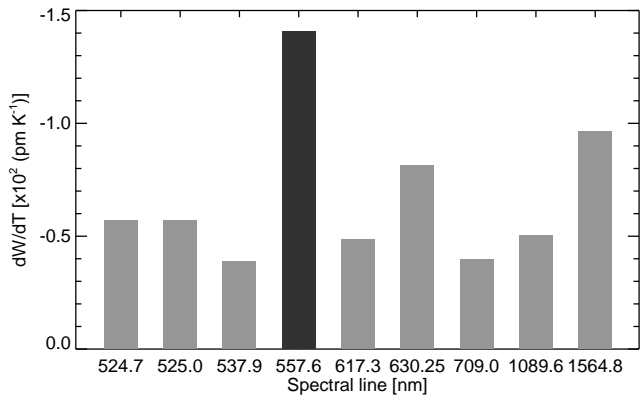}}
\end{center}
\caption{Variation of the equivalent width $W$ of 9 neutral iron 
lines when the temperature is increased by 1\,K in all photospheric
layers. The calculations have been carried out using the thermal 
stratification of the penumbral model of TTR. 
\label{rf_w} }
\end{figure}

Figure~\ref{rf_w} shows the temperature dependence of the 
equivalent width for a set of neutral iron lines commonly
used in solar physics (adapted from Cabrera Solana et al.\ 2005). 
It is apparent from this figure that \ion{Fe}{i}~557.6 nm is one of the most
temperature sensitive lines. Indeed, it shows a larger sensitivity than lines
presumed to be appropriate for temperature measurements such as \ion{Fe}{i}
524.7 and 525.0 nm. The large variation of the equivalent width of
\ion{Fe}{i}~557.6 nm with temperature is consistent with the observed behavior
of the line in the quiet sun, umbra and penumbra (Fig.~5 of Paper I).

Let us now turn our attention to the sensitivity of \ion{Fe}{i}~557.6 nm to
line-of-sight (LOS) velocities. The bottom panel of Fig.~\ref{rfs} shows the
RF of the intensity to velocity perturbations as evaluated in the penumbral
model of TTR.  These RFs correspond to a model where all velocities have been
set to zero, so they are symmetric about the line center (models in which the
LOS velocity varies with height would result in asymmetrical RFs and slightly
different sensitivities in the different atmospheric layers). As can be seen,
the line is sensitive to velocities in a height range from about $\log
\tau_{500} = 0$ to $-3.5$. The maximum sensitivity is attained in the mid
photosphere. In the deep and high layers where the continuum and the line core
are formed, respectively, the sensitivity to velocities is much reduced. The
reason has been explained by Cabrera Solana et al.\ (2005): at a given
wavelength, the intensity variation induced by Doppler shifts is proportional
to the {\em slope} of the line at that position. Thus, the intensity changes
are maximum near the line wings (which sample the mid photosphere) and minimum
near the continuum and the line core, where the slope of the intensity profile
(${\rm d}I/{\rm d} \lambda$) quickly goes to zero. The lower panel of
Fig.~\ref{rfs} also demonstrates that the intensity observed at a fixed
wavelength is sensitive to LOS velocities in a broad range of optical depths.
That is, the Doppler shifts measured from line bisectors at a given intensity
level contain information from a significant fraction of the atmosphere. Thus,
it is difficult to ascribe them to any particular layer. This limitation of
bisector analyses has long been recognized (Maltby 1964; Rimmele 1995).

\subsection{Inversion procedure}
\label{inversion}
The intensity profiles observed with TESOS have been inverted in order
to determine the stratification of temperature, LOS velocity, and
microturbulence with optical depth. To this end, we have used the SIR
code (Ruiz Cobo \& del Toro Iniesta 1992).  Despite the
high spectral resolution of our observations, the number of data
points does not allow us to use complex atmospheric models. Hence, 
we have adopted a simple one-component model with gradients of the
physical parameters. 

We derive the temperature stratification by modifying the thermal
structure of the penumbral model of TTR with two nodes. For a better
recovery of this parameter, the observed profiles are normalized to
the continuum intensity of the average quiet sun profile. Such an
absolute normalization implies that the continuum of the individual
profiles carries information about the temperature in the deep
atmospheric layers. The two nodes used to infer the thermal structure
allow for changes in the gradient and absolute position [e.g.,
$T(\tau_{500}=1)$] of the initial temperature stratification. We
assume local thermodynamical equilibrium (LTE).  Since non-LTE effects
may be important in the upper layers, our temperatures are reliable
only up to, say, $\log \tau_{500} \sim -3$.  Electron pressures are
computed from the inferred temperatures using the equation of
hydrostatic equilibrium and the ideal gas law.

The inversion also yields the stratification with depth of the LOS velocity
and microturbulent velocity. In both cases we assume linear variations with
$\log \tau_{500}$, i.e., two nodes for each parameter. In the case of the LOS
velocity, this choice seems to be appropriate considering that most observed
bisectors possess quite linear shapes (Paper II). From the inversion we also
determine the (height-independent) macroturbulence needed to reproduce the
observed line widths.  The synthetic profiles are convolved with the
instrumental profile of TESOS before being compared with the observed ones. To
describe the broadening of the line due to collisions with neutral hydrogen
atoms we use the quantum mechanical formulation of Anstee, Barklem, \& O'Mara
(see Barklem et al.\ 2000). The total number of free parameters is 7, which
compares favorably with the number of observables (100).

Without additional spectral lines observed, or measurements of the
four Stokes parameters, it is not possible to estimate the amount of
stray light contamination. We expect, however, that stray light
arising from seeing fluctuations is much reduced with respect to
earlier observations that did not benefit from adaptive optics.  By
contrast, parasitic light inside TESOS has been corrected
for. In Paper~I we determined that $\sim 5\%$ of the mean continuum
intensity contributes to each filtergram in the form of ghosts and
scattered light. This parasitic light has been subtracted from the
individual images before extracting the intensity profiles.

Our inversion approach avoids important shortcomings of simpler
analyses.  Since the complete radiative transfer problem is solved, we
effectively separate the contribution of temperatures and LOS velocities
to the observed profiles using the information provided by the
RFs. From Fig.~\ref{rfs} it is clear that the intensity at a fixed
wavelength depends on the temperature and LOS velocity stratification
along the atmosphere.  Thus, we determine both parameters
simultaneously in order not to misinterpret thermal effects as
velocity effects and vice versa. Another important improvement is that
we can ascribe the inferred velocities and temperatures to specific
optical depths, because our procedure takes into account the finite
width of the intensity contribution functions.

\begin{figure}
\begin{center}
\resizebox{.49\hsize}{!}{\includegraphics{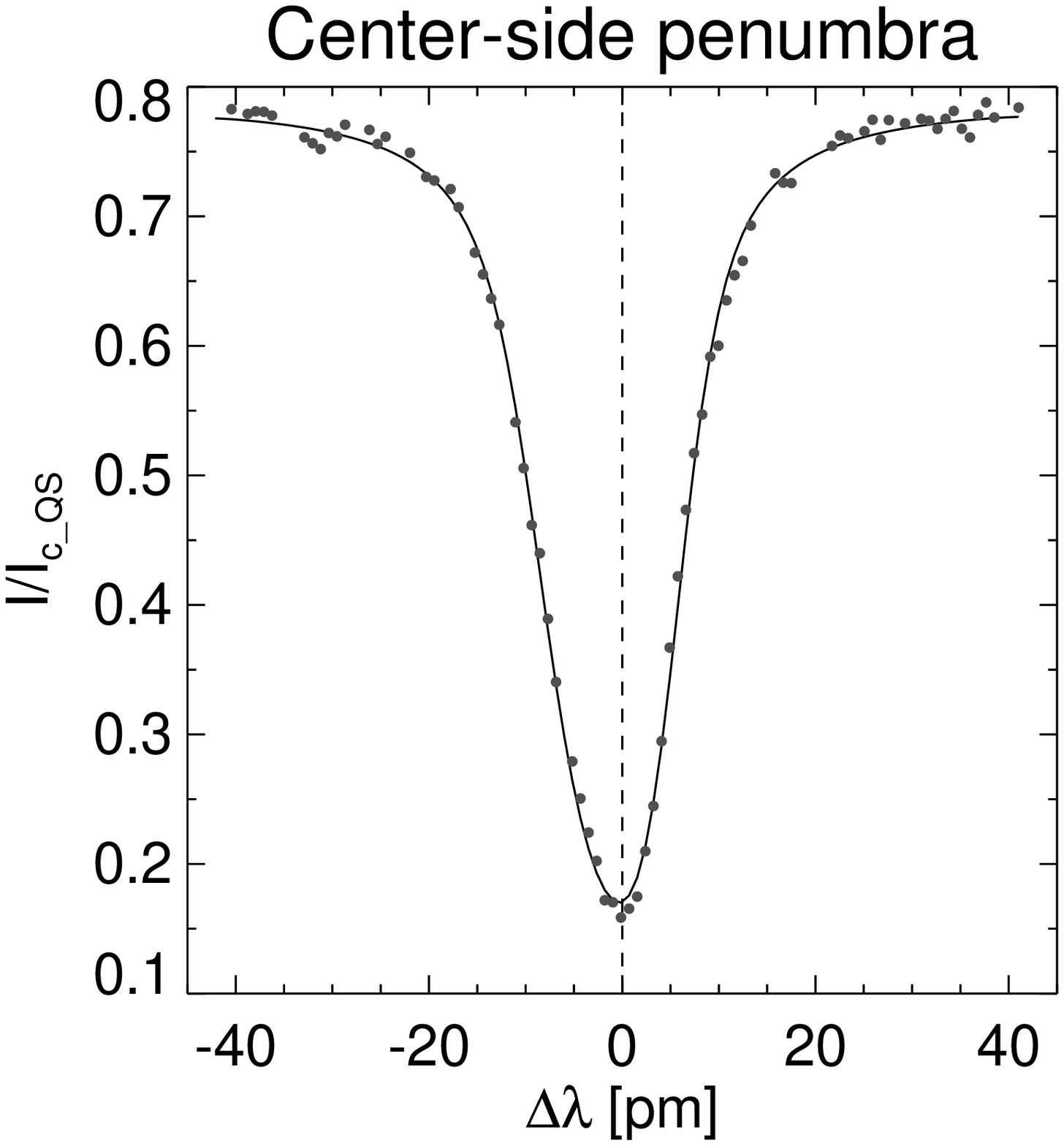}}
\resizebox{.49\hsize}{!}{\includegraphics{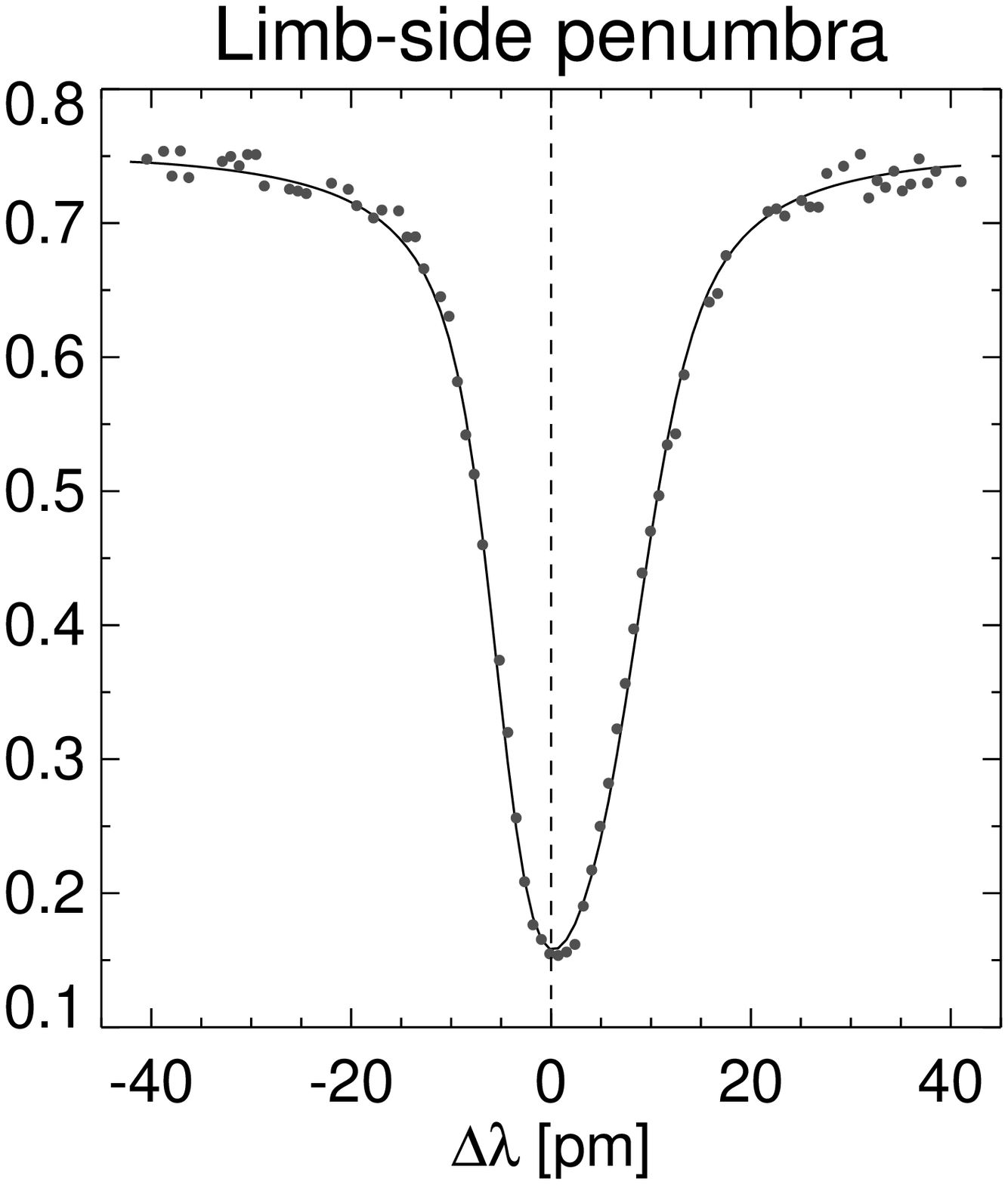}}
\end{center}
\caption{Examples of observed (dots) and best-fit (solid lines)
intensity profiles emerging from the center-side {\em (left)} and limb-side
{\em (right)} penumbra. The vertical dashed lines indicate the position of
zero velocities (see Paper I).  Note the strong asymmetries of the
line, with an extended blue wing in the center-side penumbra and 
an extended red wing in the limb-side penumbra.
\label{fit}}
\end{figure}

The simple one-component model atmosphere used in this investigation has
several limitations. First, it does not allow for any unresolved structure in
the pixel. Although our spatial resolution is one of the highest ever reached
in spectroscopic studies, we are still far from resolving the individual
constituents of the penumbra. Spectropolarimetric measurements suggest that at
least two magnetic components are necessary to understand the penumbra
(Schlichenmaier \& Collados 2002; Bellot Rubio et al.\ 2003, 2004; Borrero et
al.\ 2004, 2005). One of the two components carries most of the Evershed flow,
while the other is essentially at rest. If the flow channels are horizontally
unresolved and we do not account for this possibility, the LOS velocities
determined from the inversion would represent {\em lower} limits to the real
velocities. Second, we have chosen a very simple functional dependence to
describe the run of the LOS velocity with height. In an uncombed penumbra
consisting of penumbral flux tubes embedded in a background magnetic
atmosphere, lines of sight that pierce the tubes would 'see' discontinuities
in the LOS velocity as they encounter the upper and lower boundaries of the
tubes. Our assumption of linear velocity stratifications would provide only a
very rough description of such discontinuous velocity stratifications.  In
that case, it would be the magnitude and sign of the LOS velocity gradient
inferred from the inversion which would inform us about the velocity inside
the flux tubes and the height position of the flow channels.

Despite these shortcomings, the adopted model atmosphere does an excellent job
in explaining the observed intensity profiles.  On average, the residuals of
the fit are only slightly larger than the noise (after $2 \times 2$ binning,
the observed profiles have a signal-to-noise ratio of about 120 in the
continuum intensity). Figure~\ref{fit} shows examples of profiles recorded in
the center- and limb-side penumbra, along with the best-fit profiles resulting
from the inversion. The very asymmetrical shapes induced by the photospheric
Evershed flow are successfully reproduced, indicating that the assumption of
linear LOS velocity stratifications suffices to explain the observed Doppler
shifts and line bisectors. This is true for all profiles showing linear
bisectors. In the very outer penumbra, bisector kinks and reversals are common
(Paper~II). We cannot reproduce these profiles so well. In these cases, it
seems that more complex LOS velocity stratifications than the one used here
would apply.

\section{Results}
\label{results}
The filtergrams have been binned by a factor of 2 before extracting
the intensity profiles. This increases the pixel size to $0\farcs18
\times 0\farcs18$, which is still sufficient to sample our
seeing-limited spatial resolution of about 0\farcs5. After binning, we
are left with 37\,000 individual profiles which are inverted
independently. The inversion has been performed using a parallelized
version of SIR in a 16-processor Linux Beowulf cluster. 
The inversion of the data cube is accomplished in about 2.5 hours.

\begin{figure*}
\begin{center}
\tabcolsep 0em
\begin{tabular}{ccc}
\resizebox{0.335\hsize}{!}{\includegraphics{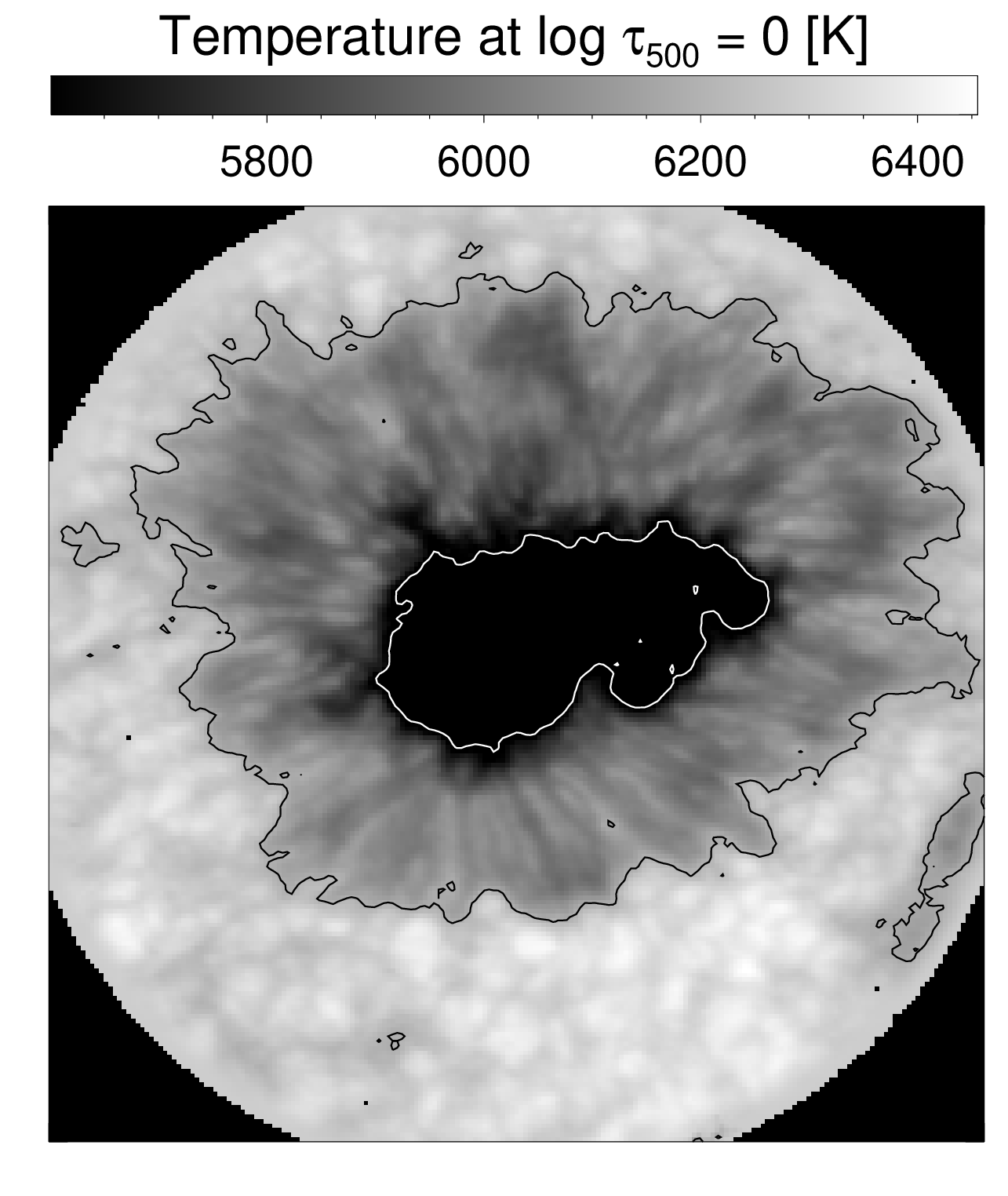}} & 
\resizebox{0.335\hsize}{!}{\includegraphics{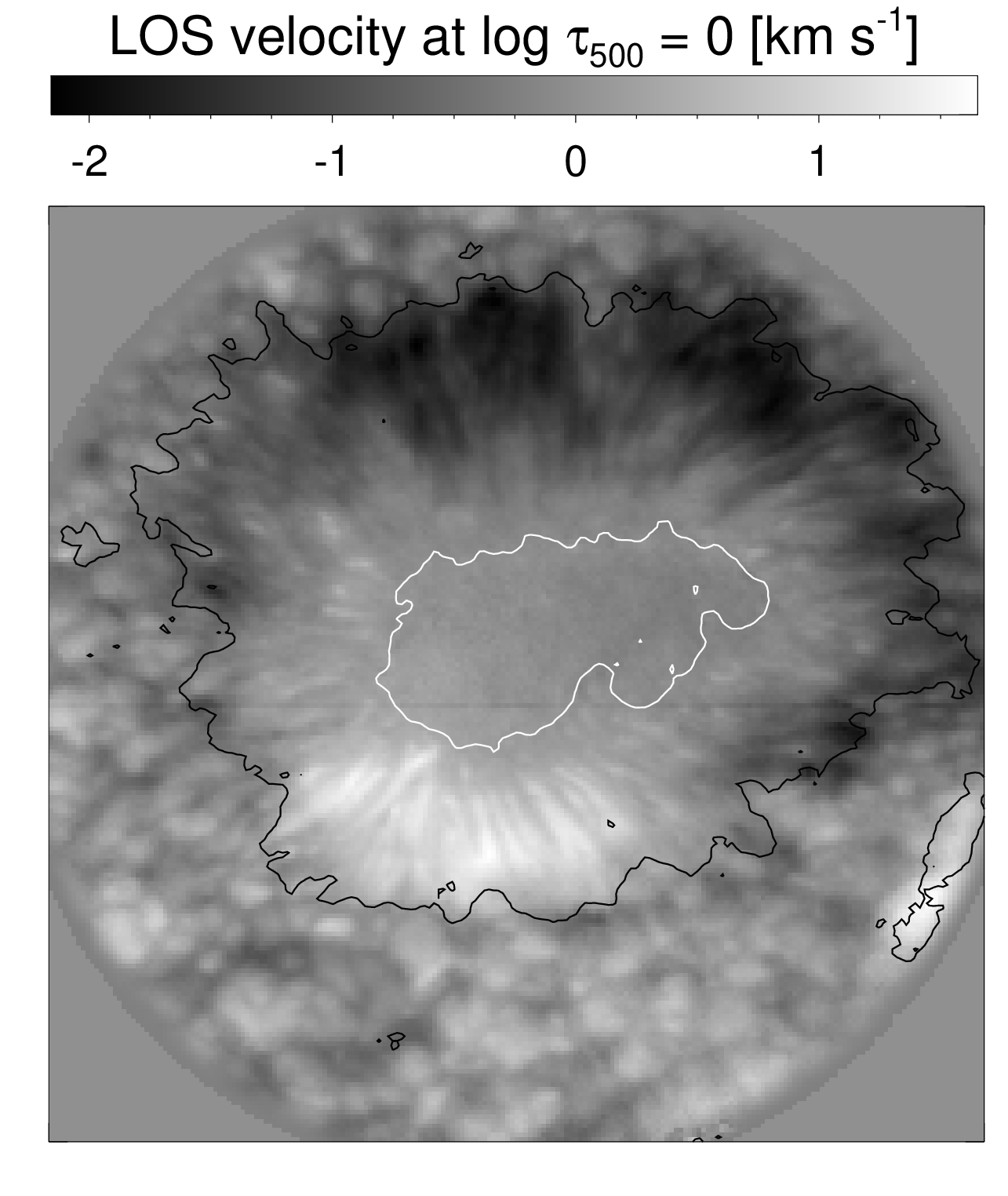}} &
\resizebox{0.335\hsize}{!}{\includegraphics{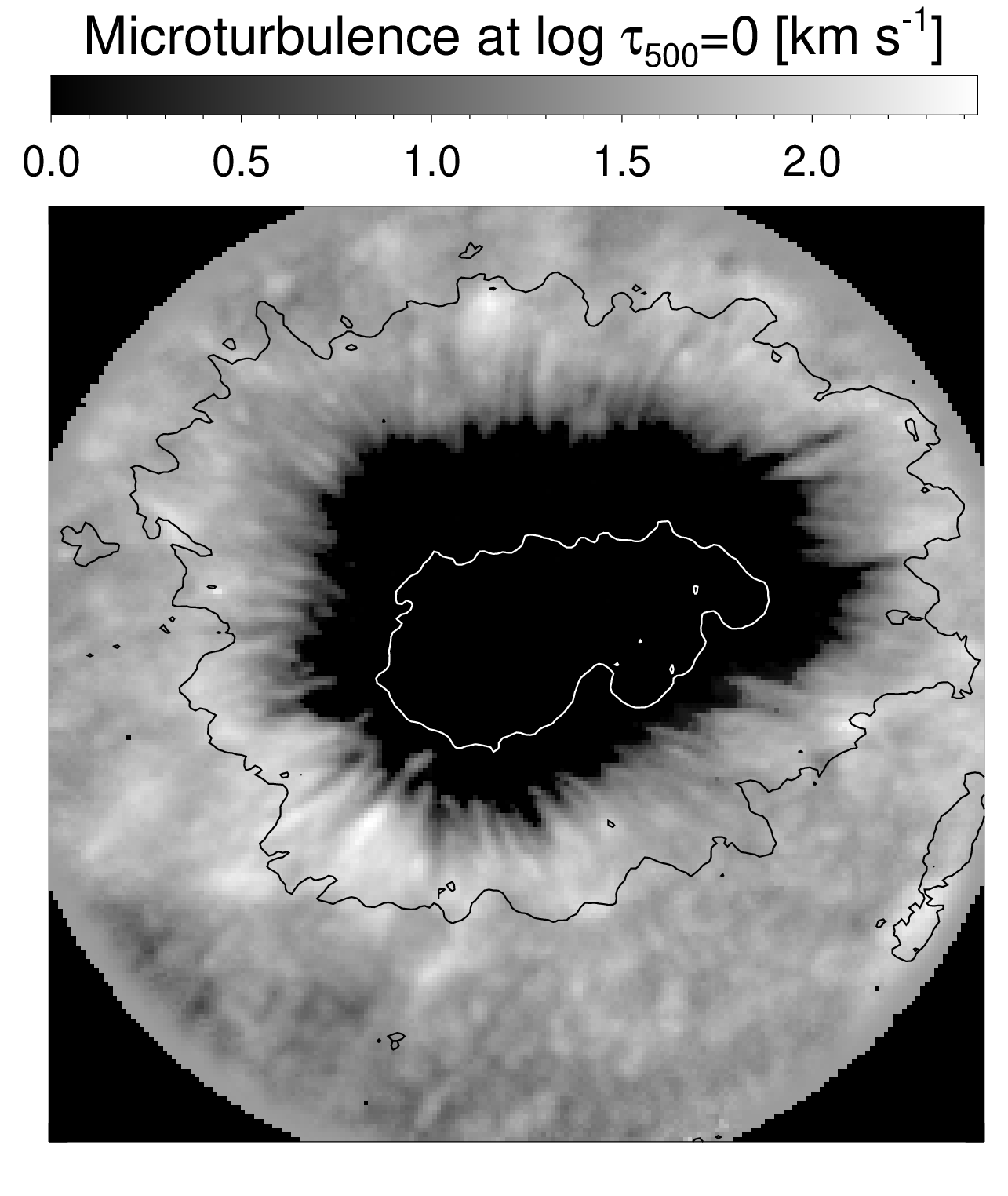}} \\

\resizebox{0.335\hsize}{!}{\includegraphics{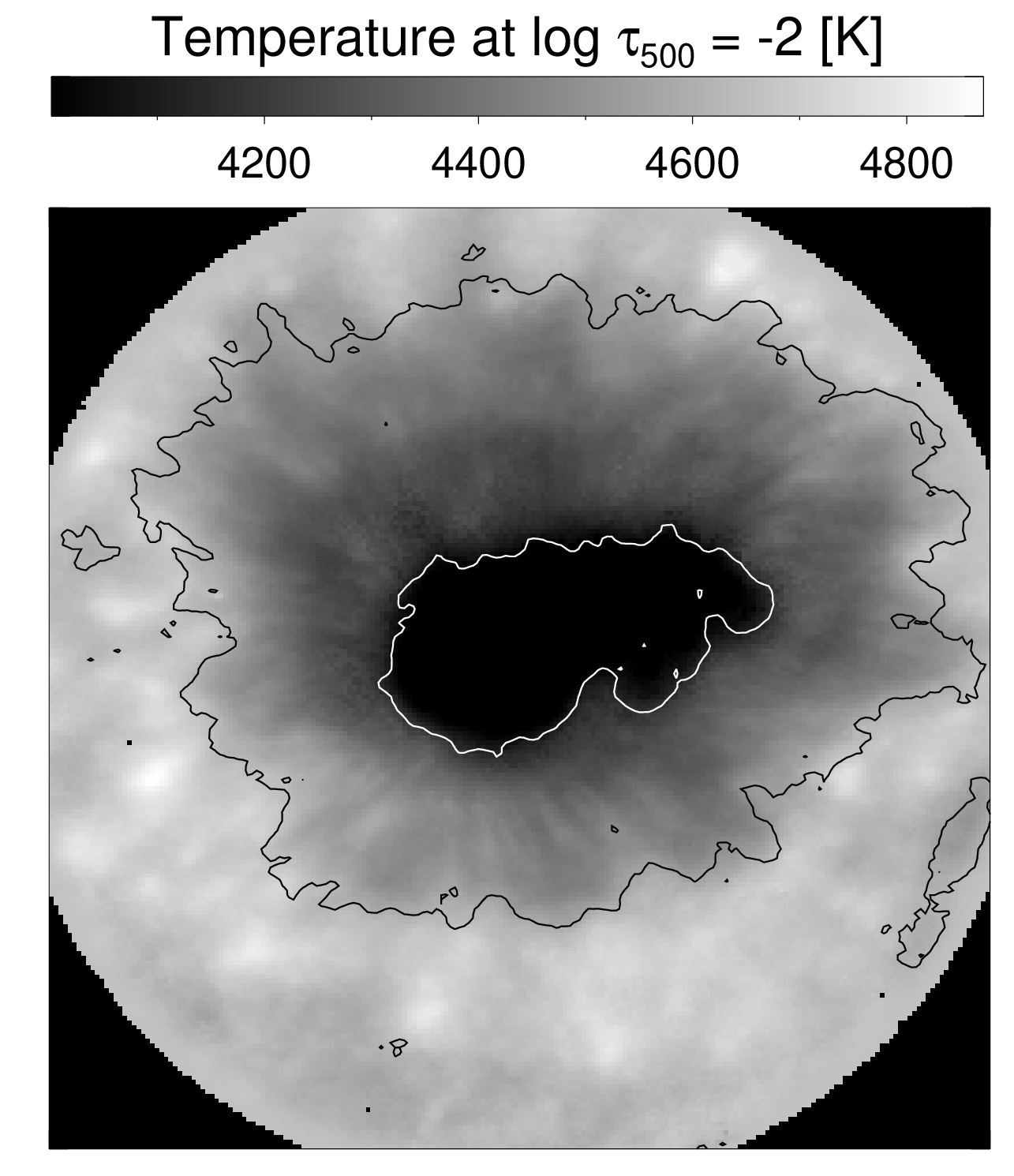}} & 
\resizebox{0.335\hsize}{!}{\includegraphics{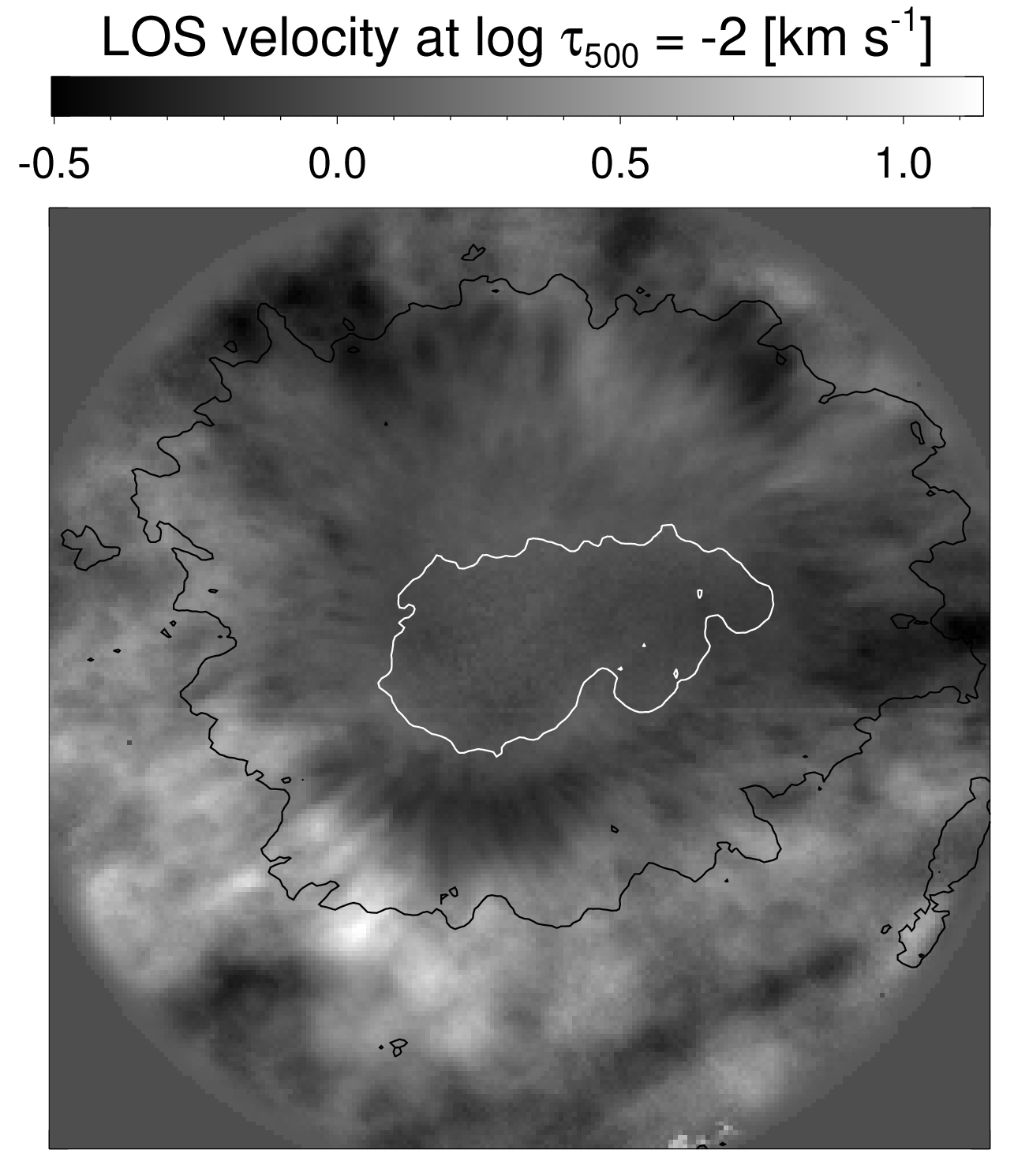}} &
\resizebox{0.335\hsize}{!}{\includegraphics{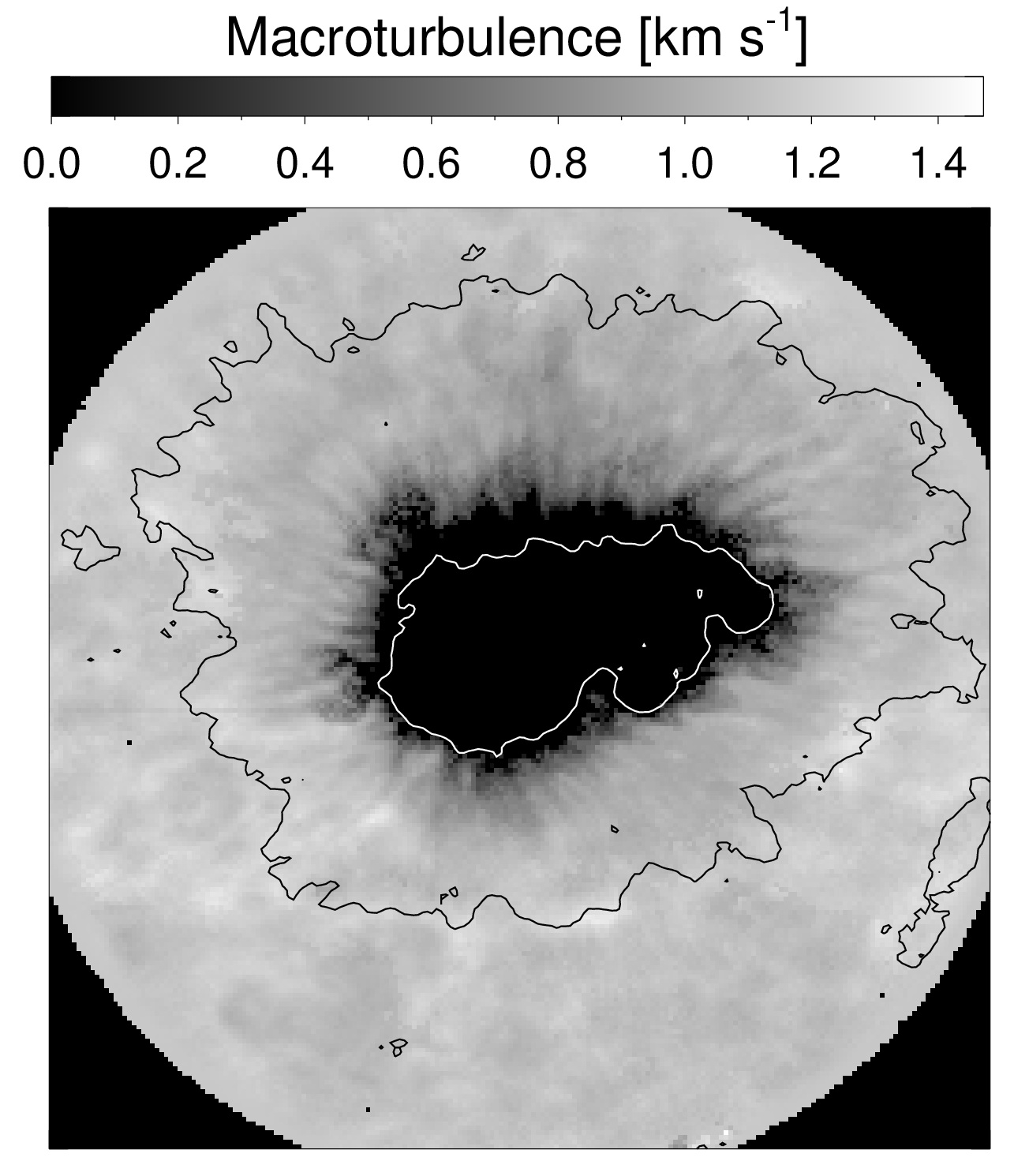}}
\end{tabular}
\end{center}
\caption{Maps of physical quantities at $\log \tau_{500}= 0$
{\em (top)} and $\log \tau_{500}= -2$ {\em (bottom)}. Depicted from left
to right are the temperature, the LOS velocity and the
microturbulence. Negative LOS velocities indicate redshifts. 
The lower right panel displays the macroturbulent velocity, 
assumed to be height independent. Contours outline the inner
and outer penumbral boundaries.
\label{maps} 
}
\end{figure*}

Figure~\ref{maps} displays maps of the temperature and LOS velocity at two
representative optical depths in the atmosphere ($\log \tau_{500} = 0$
and $-2$). The fine
structure of the penumbra is apparent (both in temperature and
velocity) in the deep photosphere, where hot penumbral filaments and
flow filaments are seen to extend from the inner to the outer
penumbral boundary. The fluctuations in temperature and LOS velocity
are much reduced at $\log \tau_{500} = -2$; indeed, no fine-scale
organization of the penumbra is obvious at this optical depth.
Consequently, the physical mechanisms producing the filaments must
operate preferentially in the deep layers.

The right panels of Fig.~\ref{maps} show maps of the microturbulence
at $\log \tau_{500} = 0$ and the macroturbulent velocity. According to 
our results, no microturbulence is needed in either the umbra or the
inner penumbra. This is in contrast to the outer penumbra, where
microturbulent velocities of 1.8--2\,km\,s$^{-1}$ are common. Thus,
the smaller line widths observed in the inner penumbra (Fig.~5 of
Paper~I) can be interpreted as being due to a reduced
microturbulence. Note that the map of microturbulence shows fine
structure in the mid penumbra. Also important is the fact that
microturbulence decreases with height, until zero values are reached
in the high layers (see Sect.~\ref{micro}). Again, this is consistent
with the idea that the mechanisms enhancing the line width operate
preferentially in deep atmospheric layers. The macroturbulence stays
re\-la\-ti\-vely constant across the penumbra, with values of 1--1.2
km\,s$^{-1}$. Only in the very inner penumbra and the umbra is the
macroturbulence reduced, perhaps as a consequence of their stronger
magnetic fields.

In the next sections, we describe in more detail the thermal and
kinematic configuration of the penumbra as deduced from the inversion.

\section{Thermal structure}
\label{thermal}

\begin{figure}
\begin{center}
\resizebox{.8\hsize}{!}{\includegraphics{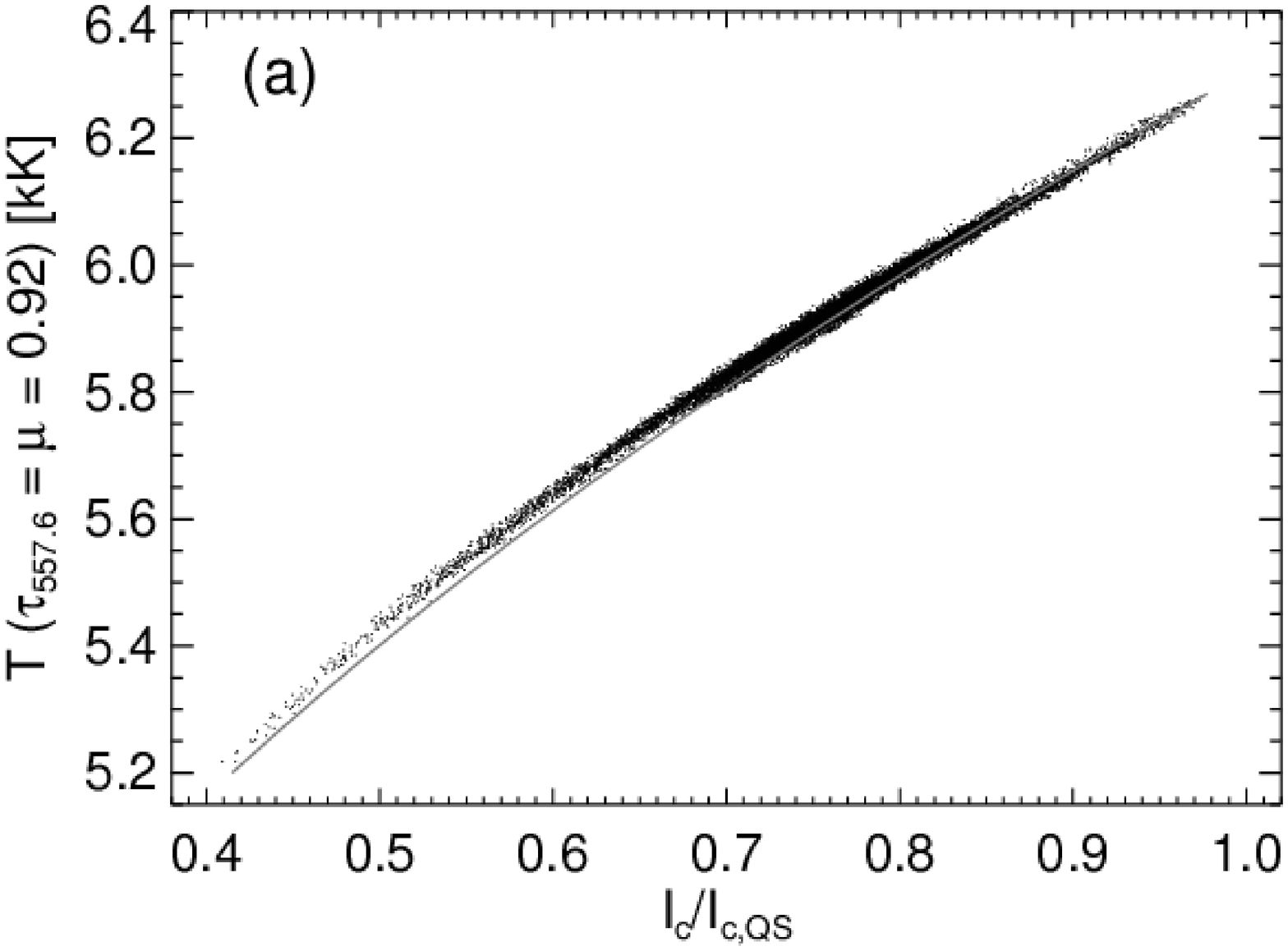}}
\resizebox{.8\hsize}{!}{\includegraphics{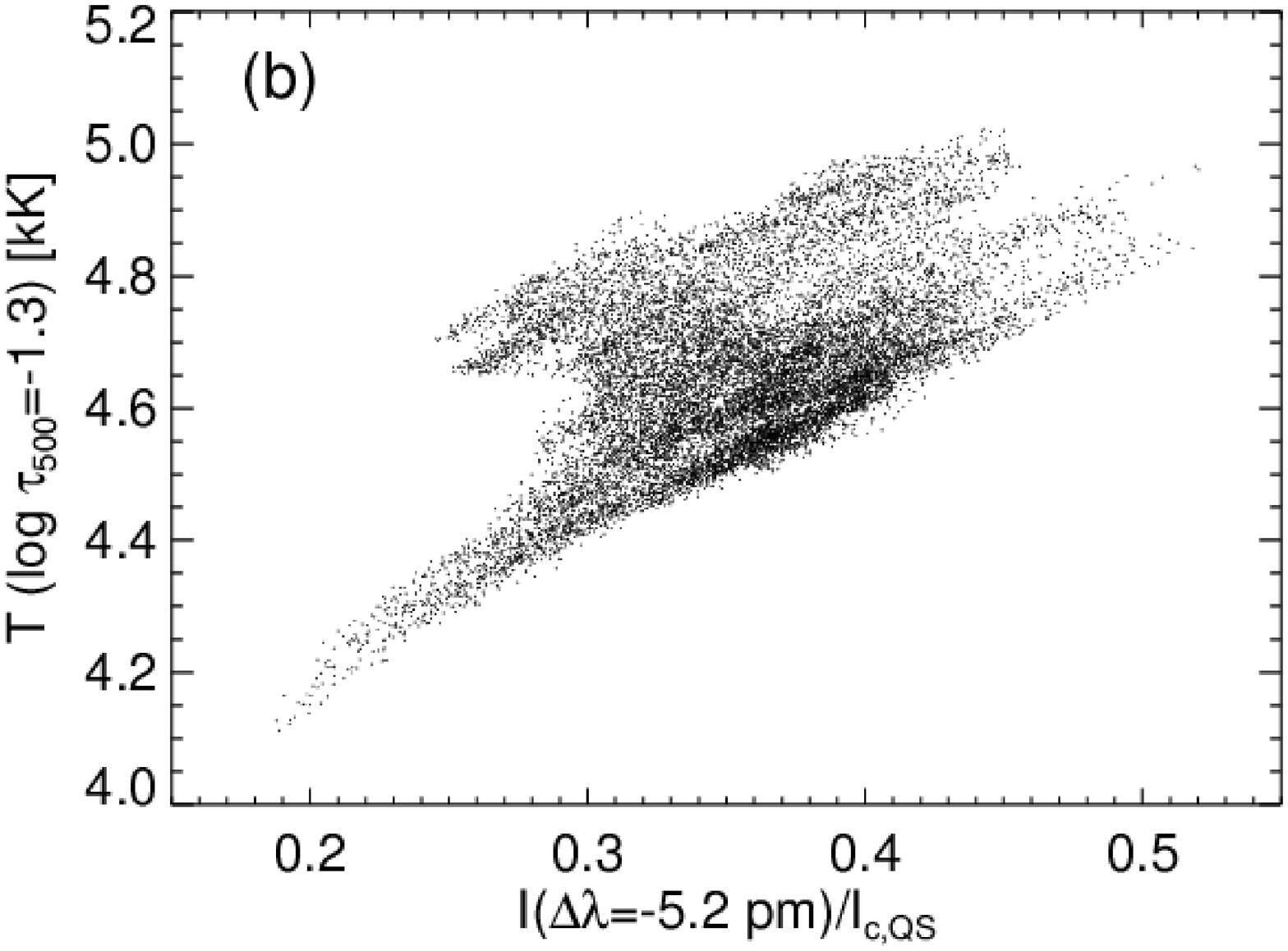}}
\end{center}
\caption{{\em (a)} Temperature at $\tau_{557.6} = \cos \theta = 0.92 $ inferred 
from the inversion vs continuum intensity (normalized to the continuum
of the average quiet sun profile). The solid line represents the 
temperatures obtained from the Eddington-Barbier 
approximation using the observed continuum intensities. {\em (b)} 
Temperature at $\log \tau_{500} = -1.3$ vs 
the intensity observed at $\Delta \lambda = - 5.2$ pm from line 
center, normalized to the continuum of the average quiet sun profile.
\label{eddington} }
\end{figure}

First, let us examine the reliability of the inferred temperatures.
Figure~\ref{eddington}a shows the temperature at $\tau_{557.6} = \cos \theta =
0.92$ returned by the inversion code versus the observed continuum
intensities. The scatter of the points is small, indicating an
excellent correlation between the two quantities. The fact that we use
an absolute normalization for the observed profiles is one of the main reasons
why the inversion code is able to determine the temperature of the
continuum-forming layers so accurately.

The solid line in Fig.~\ref{eddington}a represents the temperatures obtained
from the Eddington-Barbier approximation $I_\lambda(\theta) =
S_\lambda(\tau_\lambda = \cos \theta)$, where $I_\lambda(\theta)$ is the
specific intensity observed at an heliocentric angle $\theta$, $S_\lambda$ is
the source function (assumed to be the Planck function), and $\tau_\lambda$ is
the continuum optical depth at the wavelength of the observation. As can be
seen, the Eddington-Barbier approximation yields the same temperatures as
those deduced from the inversion. Only for small continuum intensities are the
inferred temperatures slightly larger than those indicated by the
Eddington-Barbier relation, but overall the agreement is very
satisfactory. This implies that the observed continuum intensities can be
reliably used to estimate temperatures in deep photospheric layers.  A similar
conclusion has been reached by TTR and Rouppe van der Voort (2002).

In the line wing, the one-to-one correspondence between specific 
intensity and temperature found before does not hold any longer. 
This is demonstrated in Fig.~\ref{eddington}b, where we plot the 
temperatures resulting from the inversion at $\log \tau_{500} = -1.3$ 
versus the intensities observed at $-5.2$ pm from line center.  We use
$\log \tau_{500} = -1.3$ because this is the optical depth of the centroid
of the contribution function for the line depression at $\Delta
\lambda =-5.2$ pm, evaluated in the mean penumbral model atmosphere. 
It is immediately apparent from the figure that the
same line wing intensity can be associated with a wide range of
temperatures (up to about 500 K). Thus, line wing intensities cannot
be employed to infer temperatures directly.  The reason is that the
intensities critically depend on the line asymmetries induced by the
Evershed flow and on the line width variations due to microturbulent
and macroturbulent velocities. All these factors contribute to the
large scatter of the points seen in Fig.~\ref{eddington}b. The problem
has been recognized by Balasubramaniam (2002). In an attempt to cure
it, Balasubramaniam introduced the concept of {\em flowless maps},
i.e., intensity maps constructed by removing the global Doppler shift
of the lines emerging from each individual pixel. Unfortunately, 
flowless maps are still affected by the line asymmetries caused
by the Evershed flow.

\begin{figure}
\begin{center}
\resizebox{0.78\hsize}{!}{\includegraphics{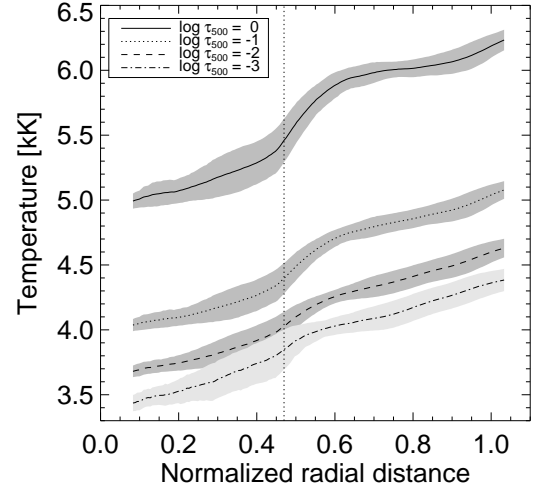}}
\vspace*{-1em}
\end{center}
\caption{Radial variation of the azimuthally averaged temperatures
at several optical depths. The shaded areas represent the standard
deviation of the temperatures at each radial distance. The vertical
line marks the inner penumbral boundary.
\label{t_rad}}
\end{figure}

Let us now investigate the radial variation of the temperature at
different optical depths. To this end, we have computed azimuthally
averaged temperatures along elliptical paths centered on the spot
(cf.\ Fig.\ 6 of Paper I). The average temperatures are plotted in
Fig.~\ref{t_rad} for four optical depths. At all heights in the
atmosphere, the temperature is observed to increase radially from the
sunspot center to the outer penumbral boundary. This increase is
rather linear except for an obvious hump in the inner penumbra, where
the temperature seems to be enhanced by several hundreds K. The
amplitude of the hump decreases as one moves towards the upper
photospheric layers.  In deep layers ($\log \tau_{500} = 0$), the hump
is so pronounced that the radial variation of the temperature is
almost flat between 0.6 and 0.9 penumbral radii (i.e., the
mid penumbra and most of the outer penumbra). We also note that the
hump is located increasingly closer to the umbra/penumbra boundary as
the upper layers are approached. These temperature enhancements could
be due to the presence of hot penumbral tubes, as suggested by the
numerical simulations of Schlichenmaier et al.\ (1998). The hot tubes
would be more easily detected in the inner penumbra because the plasma
emerging from subphotospheric layers quickly cools off by radiation
away from the tube's inner footpoint (Schlichenmaier et al.\ 1999). If
the humps exhibited by the temperature curves of Fig.~\ref{t_rad} are
indeed produced by hot flux tubes, then it is clear that they must be
located preferentially in the lower layers.

\begin{figure}
\begin{center}
\resizebox{.49\hsize}{!}{\includegraphics{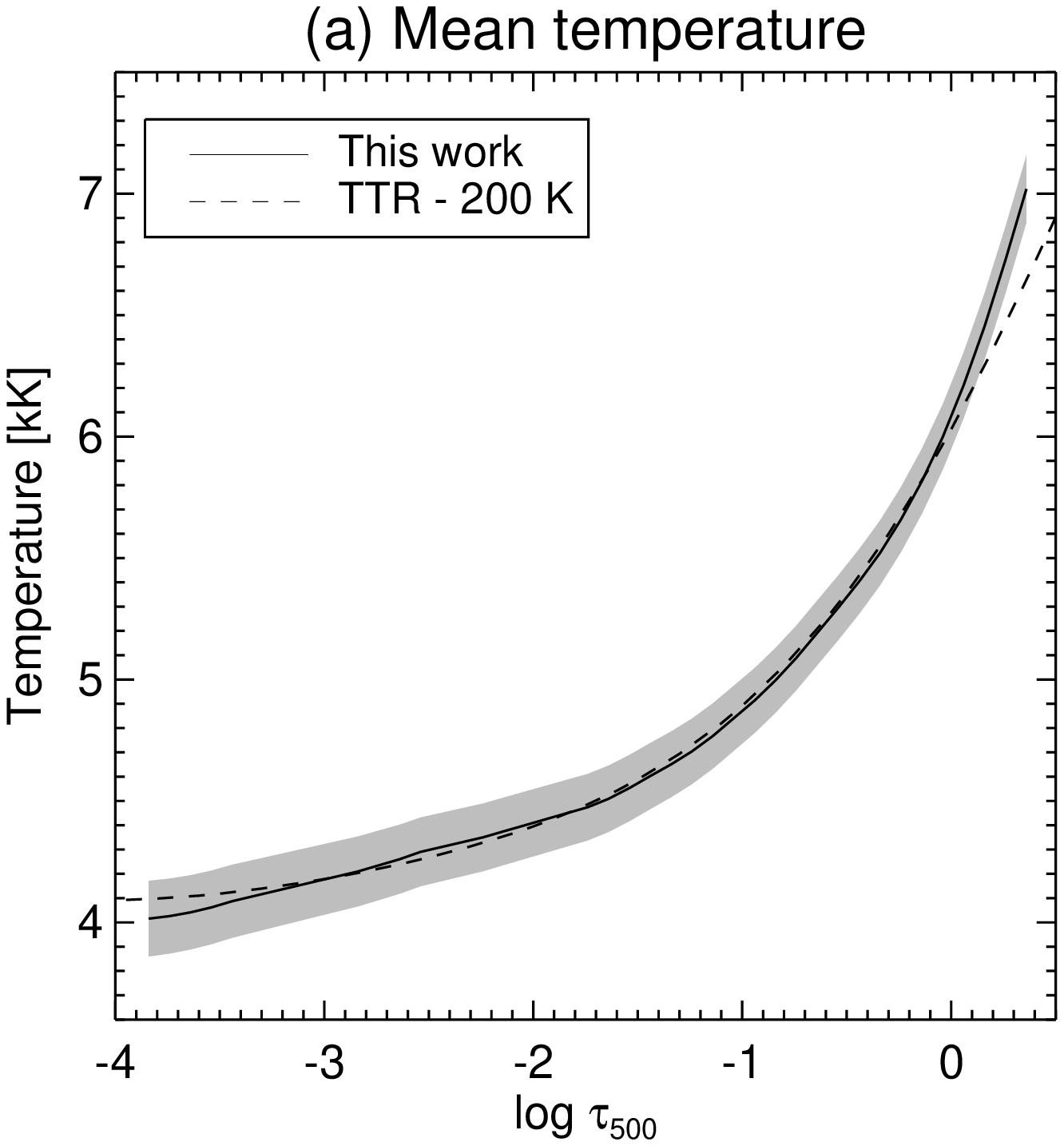}}
\resizebox{.49\hsize}{!}{\includegraphics{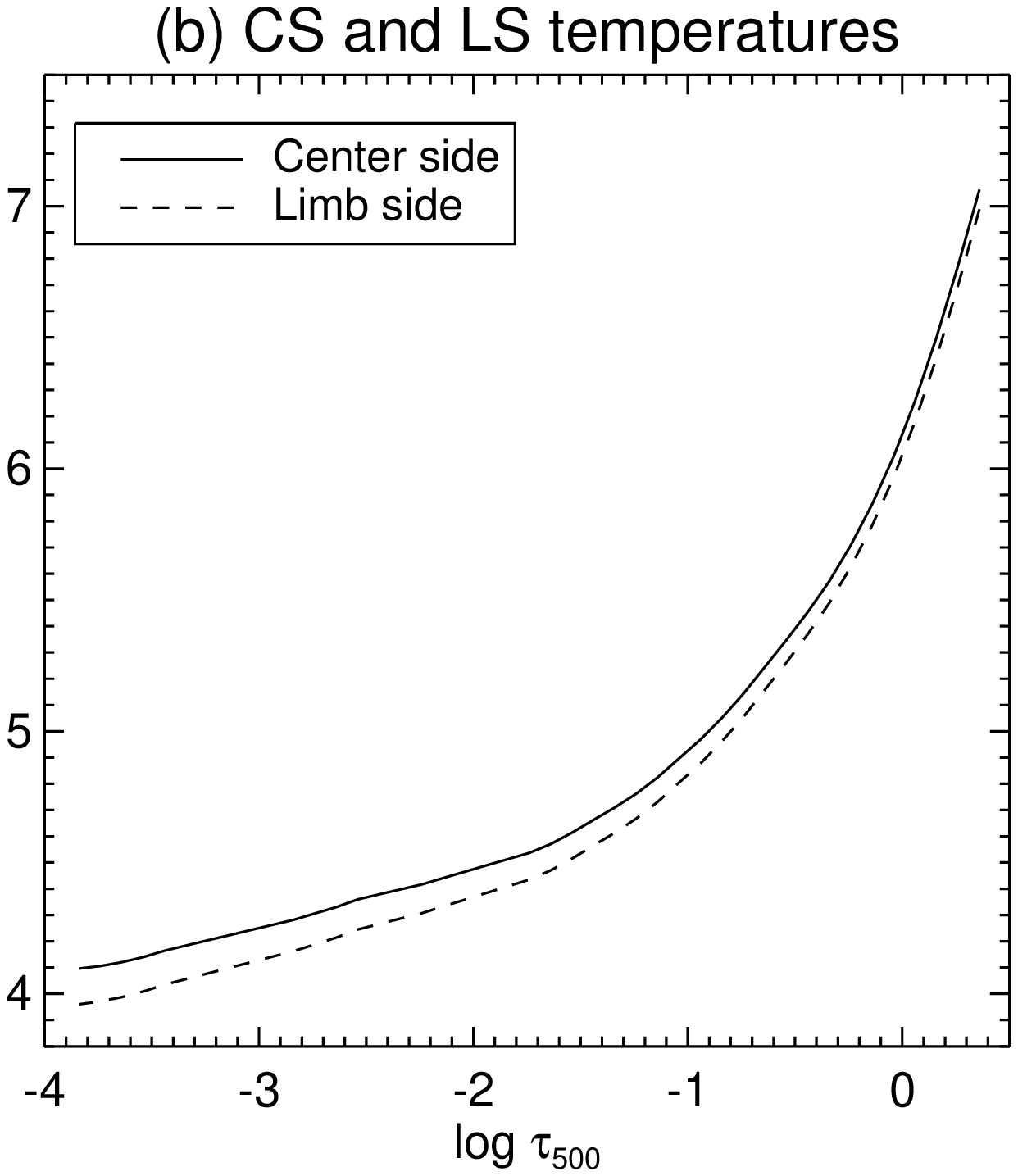}}
\end{center}
\caption{{\em (a)} Average temperature stratification in the penumbra
(solid). The penumbral model of TTR (shifted by 200 K) is
also shown for comparison (dashed). The shaded areas represent the
rms fluctuations of the temperature. {\em (b)} Average temperature
stratification in the center-side and limb-side penumbra (solid and
dashed, respectively).
\label{mean_temp}}
\end{figure}

The mean temperature stratification of the penumbra considering all pixels is
shown in Fig.~\ref{mean_temp}a. For comparison, we also plot the mean
temperature stratification obtained by TTR. The two curves have the same
curvature at all heights, but are shifted by 200~K (our temperatures being
cooler than those of TTR). We do not deem that the offset of 200~K is
significant, as different spots may certainly show different temperatures. In
addition, the two data sets may be affected by different levels of stray light
contamination.

\begin{figure}
\begin{center}
\resizebox{.71\hsize}{!}{\includegraphics{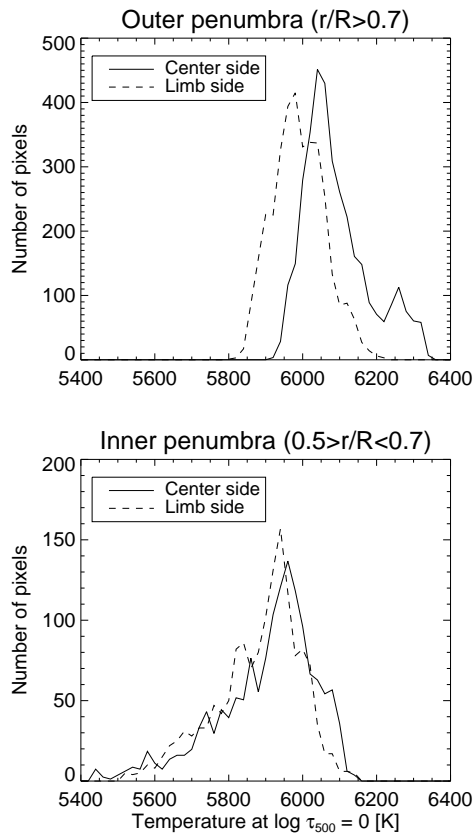}}
\vspace*{-.5em}
\end{center}
\caption{Histograms of temperature at $\log \tau_{500} = 0$ in the 
outer {\em (top)} and inner {\em (bottom)} center-side and limb-side penumbra 
(solid and dashed lines, respectively).
\label{asymmetry} }
\end{figure}

In Fig.~\ref{mean_temp}b we plot the average temperature stratifications for
the center-side and limb-side penumbra. Again, the curvature of the two
stratifications is very similar, but the limb-side penumbra is observed to be
cooler than the center-side penumbra at all heights. The difference is larger
in the upper layers (up to 150 K). A similar asymmetry has been reported by
Rouppe van der Voort (2002) from an analysis of \ion{Ca}{ii} K
observations. Rouppe van der Voort suggests that the larger average
temperature of the center-side penumbra is produced by an excess of bright
structures in the outer center-side penumbra. Our results support this
idea. In Fig.~\ref{asymmetry}, histograms of the temperature at $\log
\tau_{500} = 0$ are presented for the inner and outer parts of the center and
limb side penumbra. The inner penumbra does not show significant differences
between the two sides of the spot.  Pixels in the outer penumbra, by contrast,
are generally hotter on the center side.

The fact that both sides of the penumbra show different
temperatures might be related to the different viewing angle. Rouppe
van der Voort (2002) suggests that the higher temperatures of the
outer center-side penumbra can be explained by isotherms being tilted
downwards away from the umbra. The required tilt is opposite to what
one would expect from the Wilson depression (inclined upwards away
from the umbra, e.g., Mathew et al.\ 2004), but it may be real: it has
been demonstrated that penumbral flux tubes dive down below the
solar surface in the mid penumbra and beyond (Westendorp Plaza et
al.\ 1997, 2001; Bellot Rubio et al.\ 2004; Borrero et al.\ 2004). That is,
the field lines are slightly inclined downwards in the outer
penumbra. Perhaps this configuration of the magnetic field is capable
of determining the inclination of the isotherms, although the exact
mechanism has not been identified yet.

\begin{figure}
\begin{center}
\resizebox{.9\hsize}{!}{\includegraphics{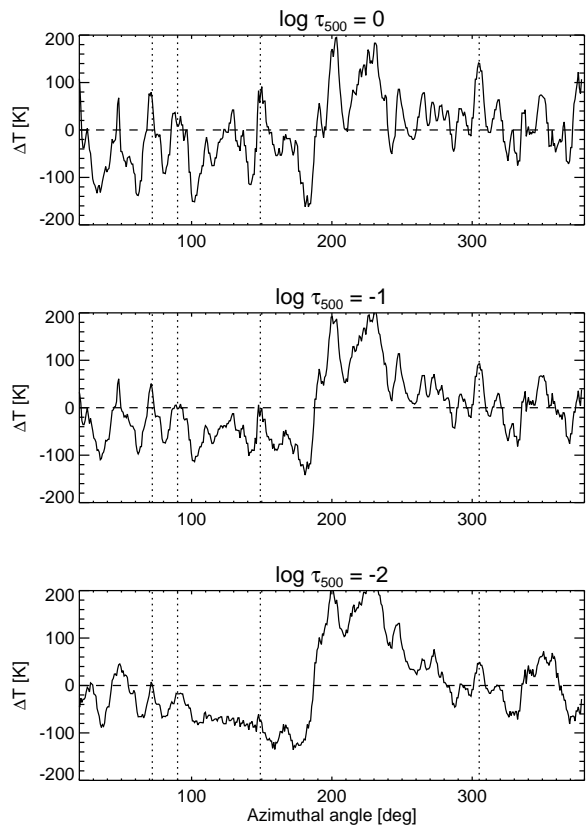}}
\end{center}
\caption{Temperature fluctuations at different optical
depths along an azimuthal path at $r/R = 0.8$. The
vertical dotted lines mark a number of selected positions 
showing enhanced temperatures. Azimuthal angles are 
measured counterclockwise, with $90^{\rm o}$ pointing to
the limb and $270^{\rm o}$ to disk center.
\label{fluctuations}}
\end{figure}

As pointed out by Rouppe van der Voort (2002), other authors also find
that the center-side penumbra is generally hotter than the limb-side
penumbra (e.g., TTR; Westendorp Plaza et al.\ 2001). To the best of our
knowledge, only one discrepant analysis exists. Schmidt \& Fritz (2004)
have studied the azimuthal variation of the continuum intensity in a
number of spots observed at different heliocentric angles. These
authors found larger intensities in the limb-side penumbra
for spots close to the solar limb, i.e., the opposite behavior. 
Clearly, more observations are required to settle the issue.

Taking advantage of the high spatial resolution of our dataset, we conclude
this section with a discussion of the temperature fluctuations observed at
different optical depths.  Figure~\ref{fluctuations} displays such
fluctuations along an azimuthal path crossing the penumbra at a normalized
radial distance of 0.8, for $\log \tau_{500} = 0$, $-1$, and $-2$. The
fluctuations become smaller and smaller as the upper layers are approached. At
$\log \tau_{500} = 0$, the temperature fluctuates by some 200~K, while at
$\log \tau_{500} = -2$ the fluctuations are smaller than 50--100~K. The
vertical lines mark a few positions where the temperature is enhanced
locally. Clearly, hot structures in the deep layers remain hot in the upper
layers, but with lower temperature contrasts.  This is in excellent agreement
with the findings of Rouppe van der Voort (2002) and Bello Gonz\'alez et al.\
(2005).  Our results suggest that the structures giving rise to the bright
penumbral filaments seen in white-light images (probably flux tubes) are
located mainly in the lower photosphere. The small temperature fluctuations
observed in high layers do not necessarily imply that flux tubes are present
there: if the deep-lying tubes are hot, it is very likely that they can heat
the layers above them. Thus, small temperature fluctuations in high layers
could be produced by tubes located deeper down.

\section{Kinematic structure}
\label{kinematic}

\begin{figure}
\begin{center}
\resizebox{0.49\hsize}{!}{\includegraphics{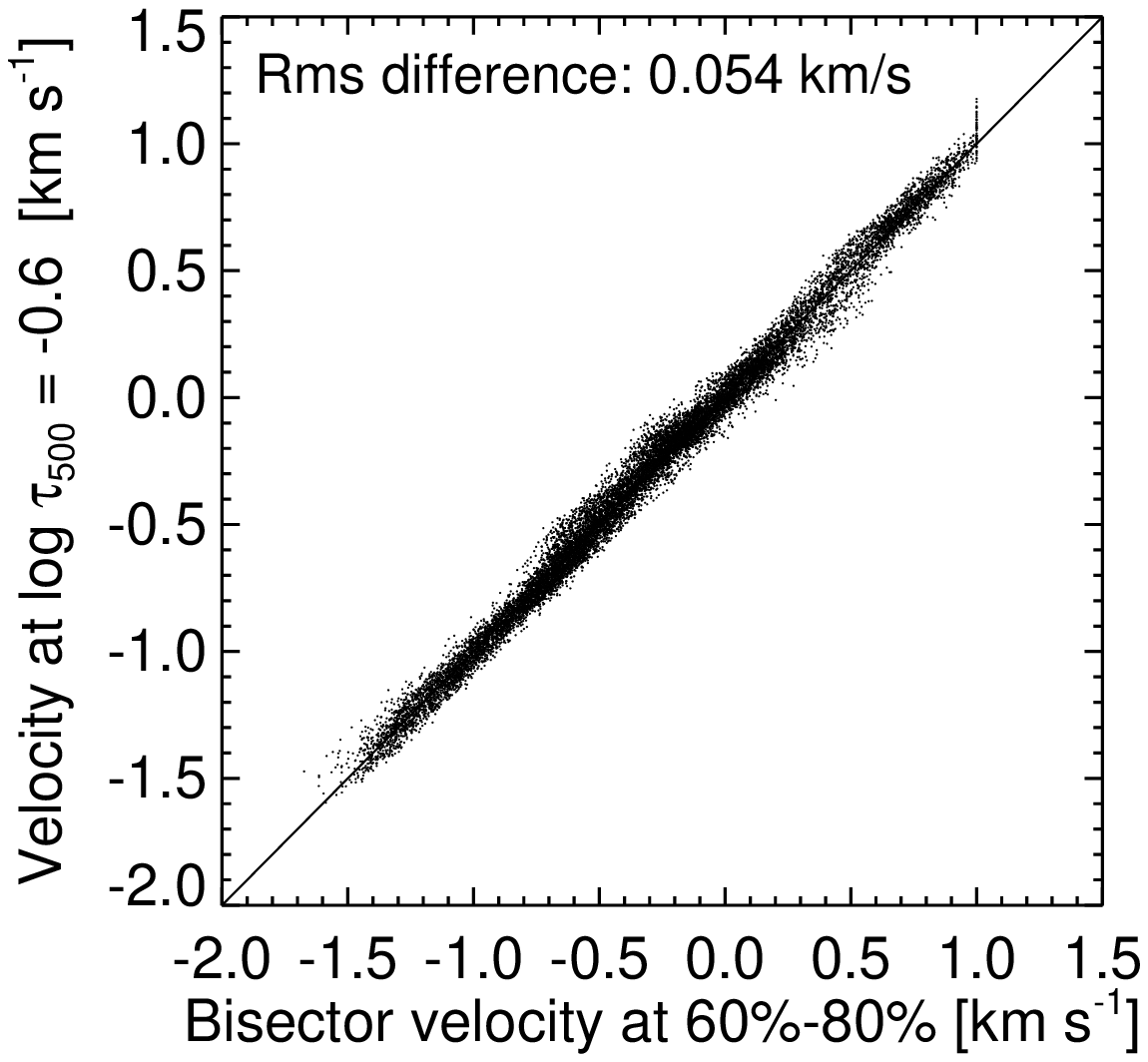}}
\resizebox{0.49\hsize}{!}{\includegraphics{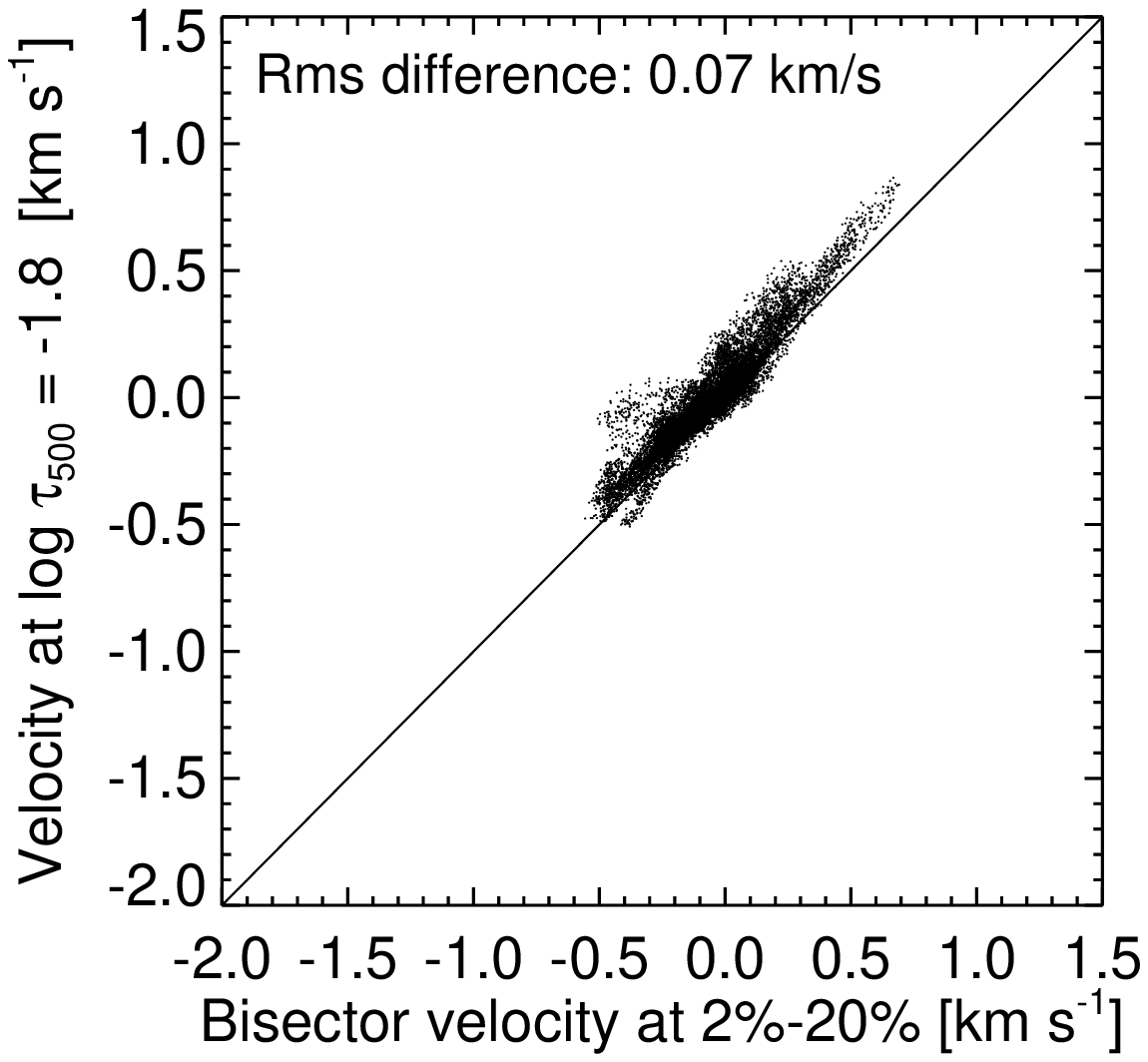}}
\end{center}
\caption{Comparison of LOS velocities resulting from the inversion and
bisector velocities at intensity levels of 60\%--80\% {\em (left)} and 
2\%--20\% {\em (right)}. The solid lines represent a one-to-one correspondence.
\label{comp_vel}}
\end{figure}

We start this section by examining how well the bisector velocities of Papers
I and II compare with the velocities inferred from the inversion. Such a
comparison is important to strengthen the conclusions drawn in Paper II from
the analysis of the observed bisectors.  Figure~\ref{comp_vel} shows scatter
plots of the LOS velocities deduced from the two different methods. As can be
seen, the agreement is very satisfactory in both deep and high
layers. Bisector velocities near the continuum (intensity levels 60\%--80\%)
are well correlated with the LOS velocities determined at $\log \tau_{500} =
-0.6$, with an rms difference of only 55\,m\,s$^{-1}$.  Bisector velocities
near the line core (intensity levels 2\%--20\%) are also well correlated with
the LOS velocities returned by the inversion at $\log \tau_{500} \sim -1.8$,
the rms difference being about 70\,m\,s$^{-1}$. We have chosen the optical
depths that yield the best agreement, but it should be remarked that they
coincide almost exactly with the centroids of the RFs to velocity
perturbations, evaluated in the average penumbral model atmosphere. The
centroids are located at $\log \tau_{500} \sim -0.8$ and 
$\sim -1.6$ for intensity levels of 60\%--80\% and 2\%--20\%, respectively. The
agreement is not surprising: Ruiz Cobo \& del Toro Iniesta (1994) and
S\'anchez Almeida et al.\ (1996) have shown that physical parameters measured
from spectral lines using simple techniques (e.g., Doppler shifts
from line bisectors) are an average of the actual stratification of the
parameter weighted by the corresponding RF. If the stratification is linear
with depth, then the height of formation of the measured parameter is the
barycenter of the RF (del Toro Iniesta 2003, Chapter 10).

Figure~\ref{comp_vel} demonstrates that Doppler velocities derived
from bisectors near the line core should be ascribed to layers
significantly deeper than the `formation height' of the line
core. This is because of the very large widths of the RFs of $I$ to
$v_{\rm LOS}$. Such widths imply that a wide range of layers
contribute to the observed Doppler shift. As shown in Fig.\ 1, the
line core reacts only little to mass motions in the upper photosphere,
and so velocities there are difficult to detect unless the
contribution of deeper layers is properly considered.

\subsection{Height variation of the LOS velocity}
\label{height_variation}

\begin{figure*}
\begin{center}
\resizebox{0.42\hsize}{!}{\includegraphics{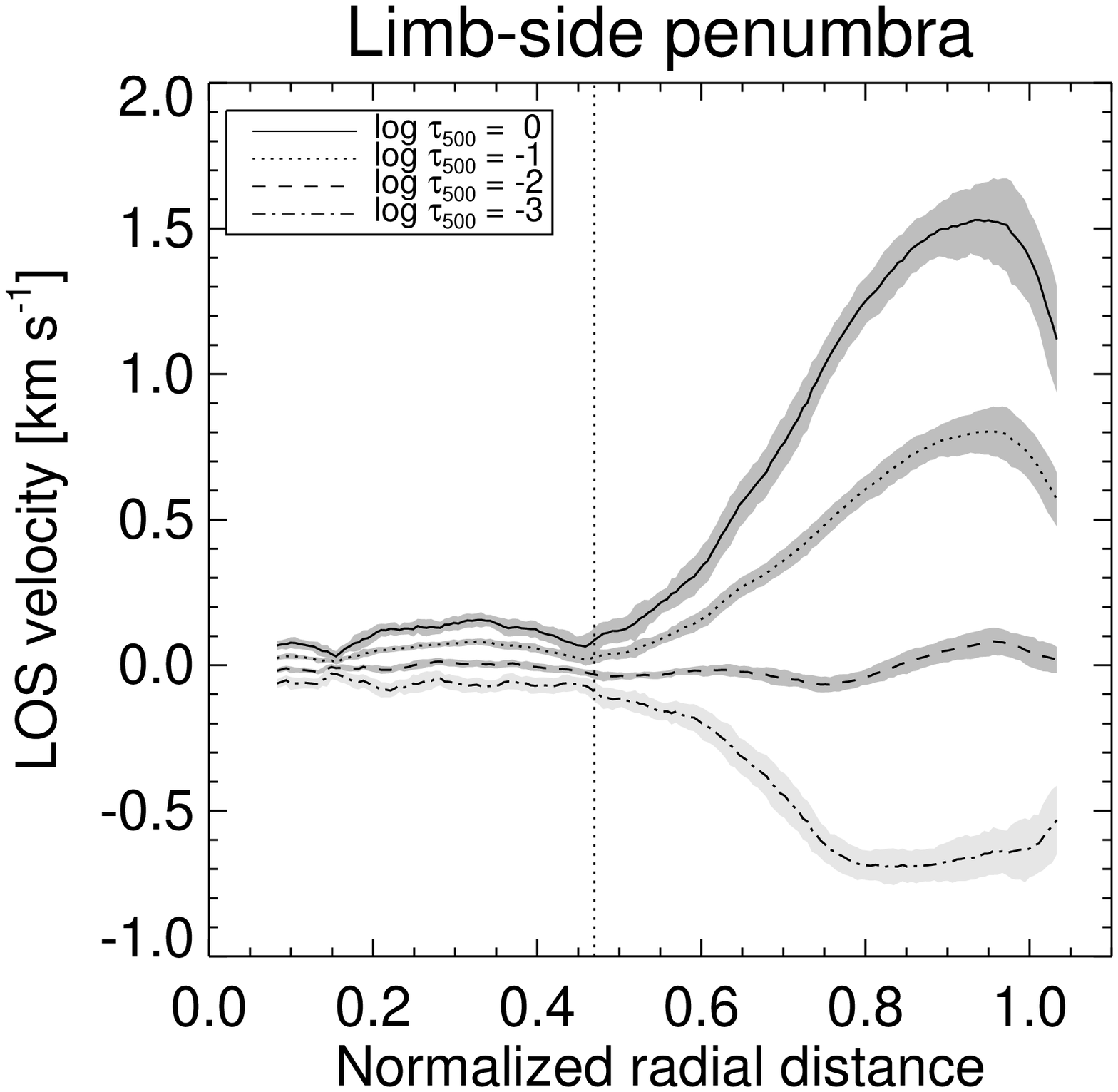}} \phantom{separa}
\resizebox{0.42\hsize}{!}{\includegraphics{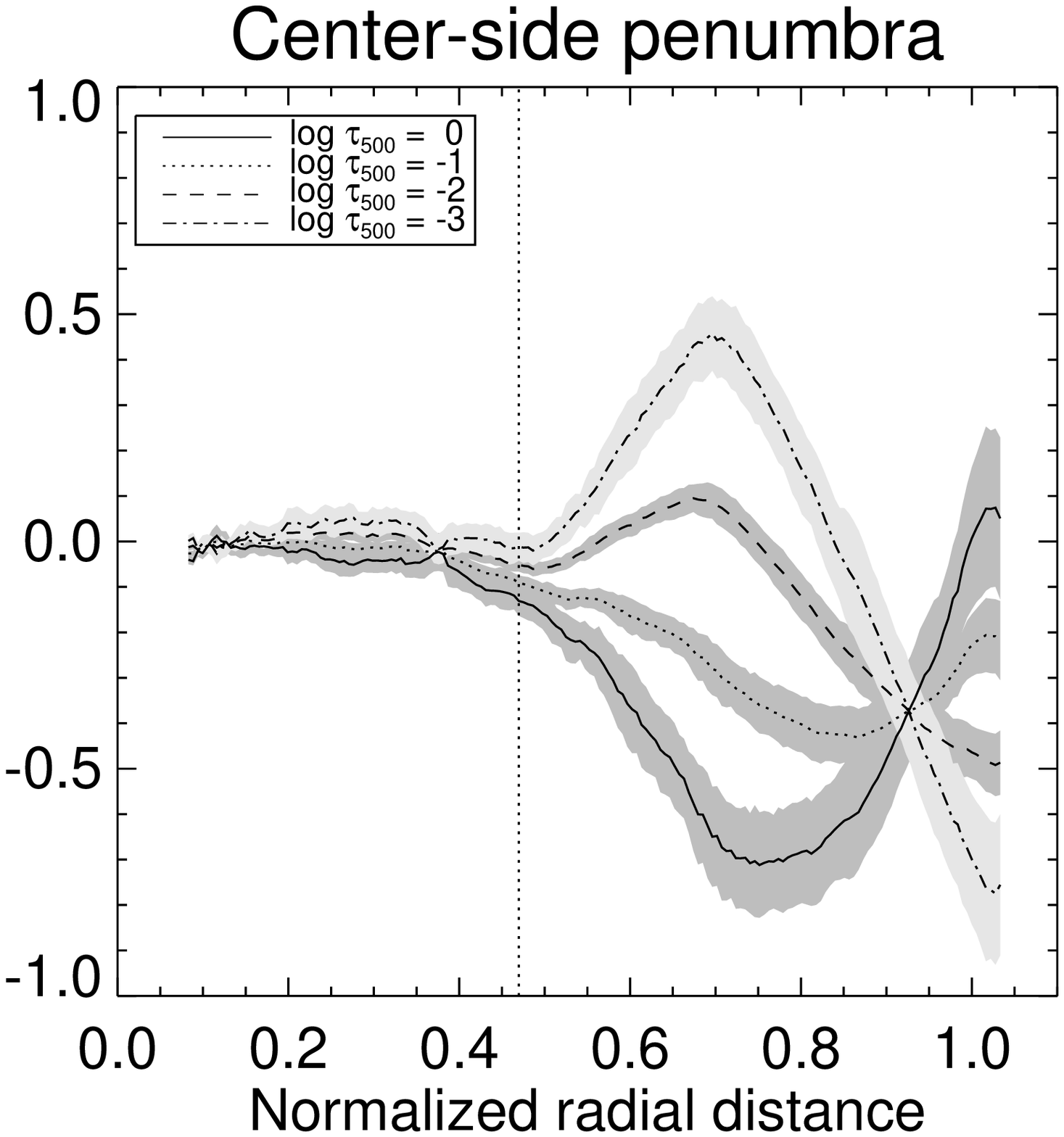}} \\
\resizebox{0.47\hsize}{!}{\includegraphics{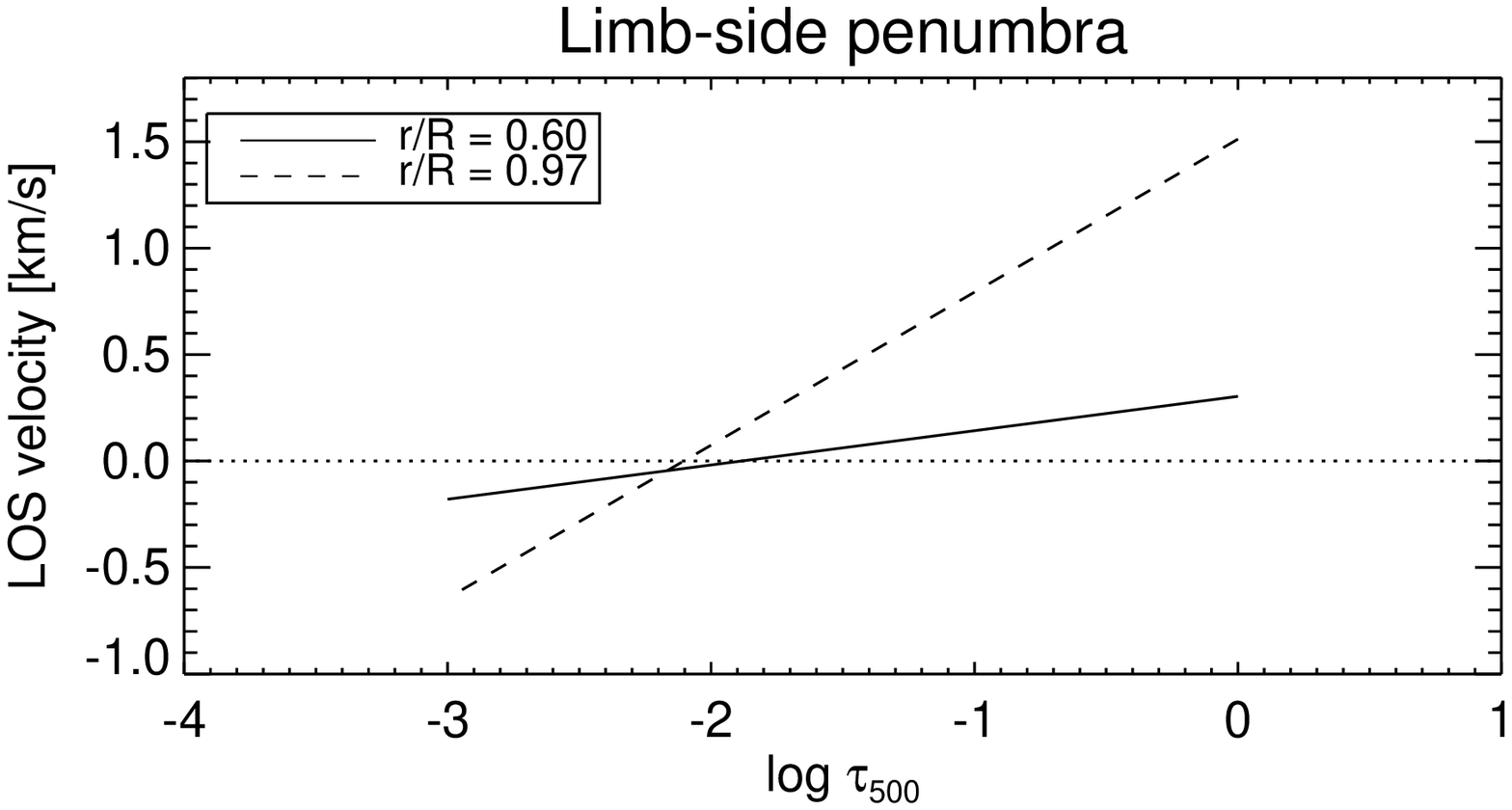}}  
\resizebox{0.47\hsize}{!}{\includegraphics{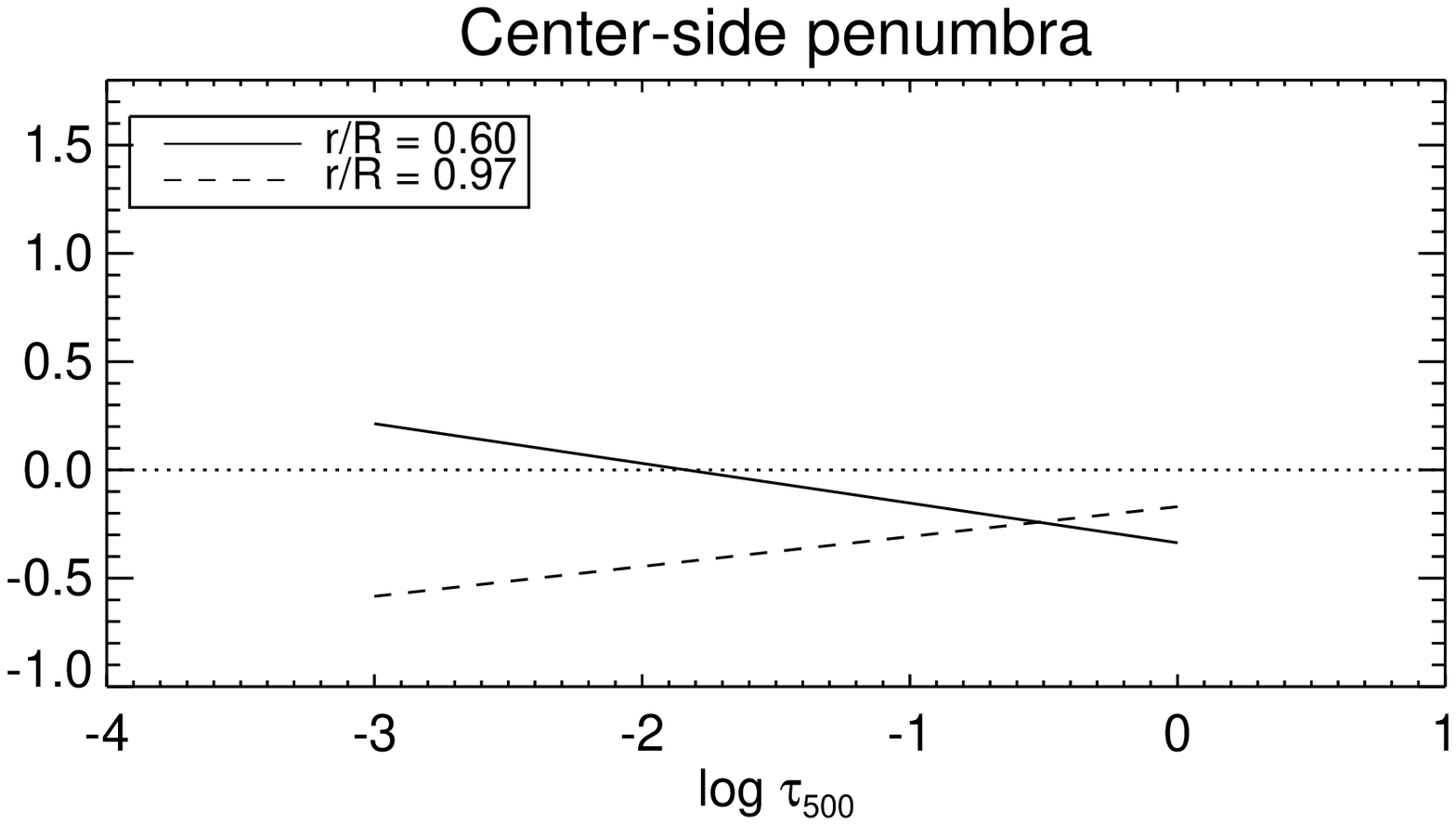}}
\vspace*{-1em}
\end{center}
\caption{{\em Top:} Radial variation of the azimuthally averaged LOS velocity
in the limb-side {\em (left)} and center-side penumbra {\em (right)}, for four
different optical depths. The shaded areas represent the rms variation
of the LOS velocities at each radial distance. Positive LOS velocities indicate
motions away from the observer (i.e., redshifts). {\em Bottom:} Average LOS 
velocity stratifications in the limb-side penumbra {\em (left)} and 
center-side penumbra {\em (right)} at $r=0.6R$ and $r=0.97R$. 
\label{averageLOS1}}
\end{figure*}

In Fig.~\ref{averageLOS1} we show the azimuthally averaged LOS velocity at
four optical depths as a function of normalized radial distance. We consider
separately the limb-side and center-side penumbra (left and right upper
panels, respectively). The curves depicted in Fig.~\ref{averageLOS1} confirm
in a more quantitative way the impression from Fig.~\ref{maps} that the
Evershed flow strongly decreases with height. A velocity reversal is observed
around $\log \tau_{500} = -2$. On the limb side, we infer redshifts in the
deep photosphere and blueshifts in the upper layers. For most of the
center-side penumbra, blueshifts in deep layers and redshifts in high layers
are retrieved (an exception is the very outer part of the center-side
penumbra). Interestingly, the LOS velocity stratifications displayed in Fig.~4
of TTR also feature sign reversals around $\log \tau_{500} = -2$. A possible
explanation for the sign reversal is offered in Sect.\ \ref{signreversal}.

The lower panels of Fig.~\ref{averageLOS1} show the average velocity
stratifications at two radial distances: $r=0.6R$ and $r=0.97R$. On the limb
side, the gradient of $v_{\rm LOS}$ with $\log \tau_{500}$ is significantly
larger in the outer penumbra. On the center side, the situation is reversed,
with slightly larger gradients in the inner penumbra.  These gradients
reproduce the slopes of the observed bisectors in the different parts of the
penumbra (Fig.~1 of Paper II).  Note the occurrence of negative velocities
(blueshifts) in the upper layers except in the inner center-side penumbra,
where positive velocities are retrieved above $\log \tau_{500} = -2$. The
inferred velocity stratifications, together with the various broadening and
smearing mechanisms, explain the line-core blueshifts observed almost
everywhere in the penumbra, as well as the small line-core redshifts detected
in the inner center-side penumbra (e.g., bisectors \#33 and \#34 in Paper
II). At this point it is important to remark that the line-core blueshifts
{\em cannot} be produced by the inverse Evershed flow. According to
Georgakilas et al.\ (2003, his Fig.\ 6), the speed and angle with respect to
the local vertical of the inverse Evershed flow are $\sim 0.5$\,km\,s$^{\rm
-1}$ and $100^{\rm o}$ in the inner penumbra, at the height where the
intensity observed at 0.5 \AA\/ from the center of the H$\alpha$ line is
formed. The flow speed and the flow angle near the outer penumbral boundary
are $\sim 4.5$\,km\,s$^{\rm -1}$ and $115^{\rm o}$, respectively.  With 
these flow configurations and the heliocentric angle of our observations, 
the inverse Evershed effect would produce non-negligible Doppler shifts only 
in the outer center-side penumbra, where it would show up as a {\em redshift},
not as a blueshift.

\subsection{Flow geometry at different heights}

\begin{figure}
\begin{center}
\resizebox{.93\hsize}{!}{\includegraphics{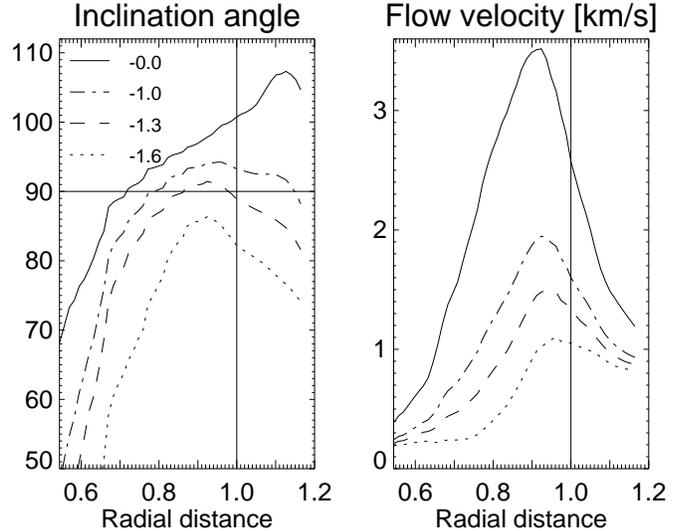}}
\end{center}
\caption{Flow geometry at different optical depths as deduced from
the azimuthal variation of the observed LOS velocity. {\em Left:} 
Flow inclination (with respect to the local vertical). {\em Right:} 
Absolute magnitude of the flow.
\label{flowgeometry} }
\end{figure}

We have computed the flow speed and flow angle at different optical depths
from the azimuthal variation of the LOS velocities returned by the inversion
code. The calculations have been performed as explained in Sect.\ 5.4 of
Paper~I. If one assumes that the flow field is axially symmetric, the
azimuthally averaged LOS velocity reflects the vertical flow component, and
the amplitude of the azimuthal variation of the LOS velocity reflects the
horizontal flow component.  Using these components, the mean flow inclination
and the absolute flow velocity can be deduced. The results of the analysis 
are presented in Fig.~\ref{flowgeometry} as a function of normalized radial 
distance.

As mentioned in Sect.~\ref{inversion}, our simple model atmosphere does not
account for the possibility of unresolved structure in the resolution
element. Thus, the flow speeds resulting from the calculations are probably
smaller than the real ones\footnote{Following Bellot Rubio (2004), we may
assume for simplicity that at a given radial distance the flow channels occupy
a fraction $0 \leq \alpha \leq 1$ of the resolution element, the remaining
$(1-\alpha)$ being plasma at rest. As a rough estimate, the LOS velocity
observed in the resolution element would be $v_{\rm LOS} \sim \alpha
\tilde{v}_{\rm LOS}$, with $\tilde{v}_{\rm LOS}$ the intrinsic LOS velocity of
the flow channel.}. To first order, however, the flow angles are not
influenced by this effect, because they are derived from the {\em ratio} of
the horizontal and vertical components of the flow, and both are affected
equally by the lack of spatial resolution.

From Fig.~\ref{flowgeometry} we note the following features:
\begin{enumerate}
\item The flow inclination (with respect to the local vertical) increases with
optical depth. Downflows are present in the mid and outer penumbra, but only
in deep layers. In high layers, the flow is always inclined upward.
\item The flow velocity decreases with decreasing optical depth.
\item The radial position of maximum flow velocity and maximum 
flow inclination 'migrate' outwards as we go to higher atmospheric 
layers.
\end{enumerate}

In summary, the {\em azimuthally averaged} flow velocity increases with depth
and the {\em azimuthally averaged} flow inclination show downflows only in the
deep layers. This strengthens the conclusions of Paper~II, where we inferred a
similar flow configuration from the interpretation of individual bisector
shapes. It is important to remark that calculations using the bisector
velocities derived in Paper~II lead to the very same flow angles and flow
speeds (cf.\ Fig.~6 in Bellot Rubio 2004), as could have been expected from
Fig.~\ref{comp_vel}.

The results presented in Fig.~\ref{flowgeometry} are in good agreement with
the flow inclinations determined by Bellot Rubio et al.\ (2003, 2004) from a
two-component inversion of infrared Stokes profiles. The radial variation of
the flow angle described by Bellot Rubio et al.\ corresponds roughly to our
curve for $\log \tau_{500} = 0$, because the \ion{Fe}{i} 1565 nm lines they
use are sensitive to velocities in very deep and narrow photospheric
layers. Our results are also consistent with those of S\'anchez Cuberes et
al.\ (2005), who inverted spectropolarimetric measurements of a sunspot at
disk center in terms of one-component model atmospheres with gradients of
velocity. These authors found that downflows occur predominantly in deep
layers ($\log \tau_{500} < -1$), disappearing higher up in the atmosphere.
The present analysis, however, does not confirm the small downflows
retrieved by S\'anchez Cuberes et al.\ (2005) in the inner penumbra.

\subsection{Microturbulence}
\label{micro}
\begin{figure}
\begin{center}
\resizebox{0.8\hsize}{!}{\includegraphics{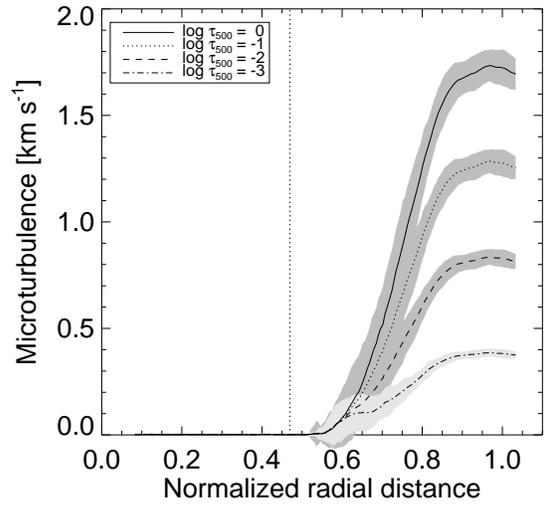}}
\end{center}
\caption{Radial variation of the azimuthally averaged microturbulence
at four optical depths. The shaded areas represent the standard deviation 
of the microturbulent velocities found at each radial distance. The 
vertical line marks the inner penumbral boundary.
\label{mic_rad}}
\end{figure}

For each pixel the inversion returns the height variation of the
microturbulence, assumed to be linear with $\log \tau_{500}$.
Figure~\ref{mic_rad} shows the azimuthally averaged microturbulence at four
optical depths versus normalized radial distance. From this figure it is
apparent that:
\begin{enumerate}
\item In the inner penumbra, up to about 0.6 penumbral radii, the
microturbulence is zero at all heights.
\item In the mid and outer penumbra, the microturbulence is different from
zero and increases with optical depth and radial distance.
\item The gradient with depth of the microturbulence increases with increasing
radial distance.
\end{enumerate}

In Paper I (Fig.~7) we determined the radial dependence of the azimuthally
averaged equivalent width (EW) and the full width at half maximum (FWHM) of
the intensity profiles observed in NOAA 10019.  For both parameters we found
minima in the inner penumbra ($r \sim 0.65 R$) and maxima in the outer
penumbra ($r \sim 0.95 R$). According to the present results, this radial
increase of the EW and the FWHM is due to a microturbulence that increases
radially (item 2 above).

What is the origin of the microturbulent velocities returned by the inversion
code? One can speculate that stronger Evershed flows result in increased shear
at the boundary layer separating the flow channels from the surrounding
(static) magnetic atmosphere. Strong shears may produce small-scale turbulence
through the onset of instabilities. This would explain why we detect enhanced
microturbulence in the mid and outer penumbra, where the flow speeds are
larger.

\begin{figure}
\begin{center}
\resizebox{.9\hsize}{!}{\includegraphics{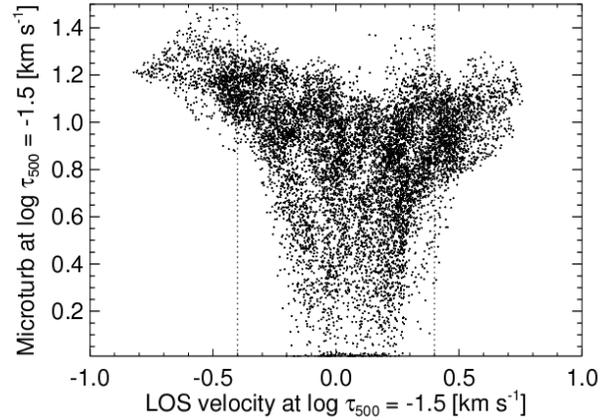}}
\end{center}
\caption{LOS velocities determined from the inversion at $\log \tau_{500}
= -1.5$ versus the microturbulence at the same optical depth. The
vertical dotted lines indicate LOS velocities of $\pm
0.4$\,km\,s$^{-1}$. Larger velocities are always associated with 
enhanced microturbulence. The same correlation is seen in deeper layers, but
less clearly.
\label{mic_vs_vel}}
\end{figure}

However, we also found in Paper~I that both the EW and the FWHM tend to be
slightly enhanced in the regions of maximum Doppler shifts (i.e., along the
line connecting the disk center and the sunspot center). A similar tendency is
observed for the microturbulence.  Figure~\ref{mic_vs_vel} shows the
microturbulence determined from the inversion at $\log \tau_{500} = -1.5$ versus
the LOS velocity at the same optical depth. From this figure, it is clear that
large LOS velocities are always associated with large microturbulences.  The
findings of Paper~I led us to conjecture that the increase in the EW and the
FWHM associated with large velocities may be due to the presence of strongly
Doppler-shifted (but spectrally unresolved) line satellites. This suggests
that part of the microturbulence we infer may not represent real small-scale
turbulence, but a convenient means to account for the effects of line
satellites.

\section{Discussion}
\label{discusion}

\subsection{Origin of enhanced line-wing intensities in the inner penumbra}
From an analysis of the Stokes profiles of the \ion{Fe}{i} 1564.8 nm line,
Bellot Rubio (2003) found that the intensity emerging from the inner penumbra
is higher than that emerging from the outer penumbra at wavelengths close to,
but not exactly at, the line center. That is, sunspots exhibit a bright ring
in the inner penumbra when observed in the wing of spectral lines. The bright
ring can be interpreted as a temperature enhancement in the mid-photospheric
layers (as it is detected in the wings of the line but not in the continuum or
line core).

Our high spatial resolution observations of NOAA 10019 confirm this behavior
(Fig.\ 8 of Paper~I) and at the same time add new information on it: the
intensity enhancement occurs in the form of radial, narrow fibrils very
similar to the bright penumbral filaments seen in continuum images, but with
much higher contrast (Fig.\ 3 of Paper~I). Moreover, the enhancement is
observed preferentially in the inner center-side penumbra blueward of the line
core, and in the inner limb-side penumbra redward of the line core. The
temperature stratifications of Fig.~\ref{t_rad} indicate that the inner
penumbra is always cooler than the outer penumbra at all heights in the
atmosphere, so the enhanced brightness cannot be due to a temperature effect.
Heating of the mid photosphere was the mechanism originally proposed by
Bellot Rubio (2003). It is also the scenario favored by S\'anchez Cuberes et
al.\ (2005).

\begin{figure}
\begin{center}
\resizebox{0.49\hsize}{!}{\includegraphics{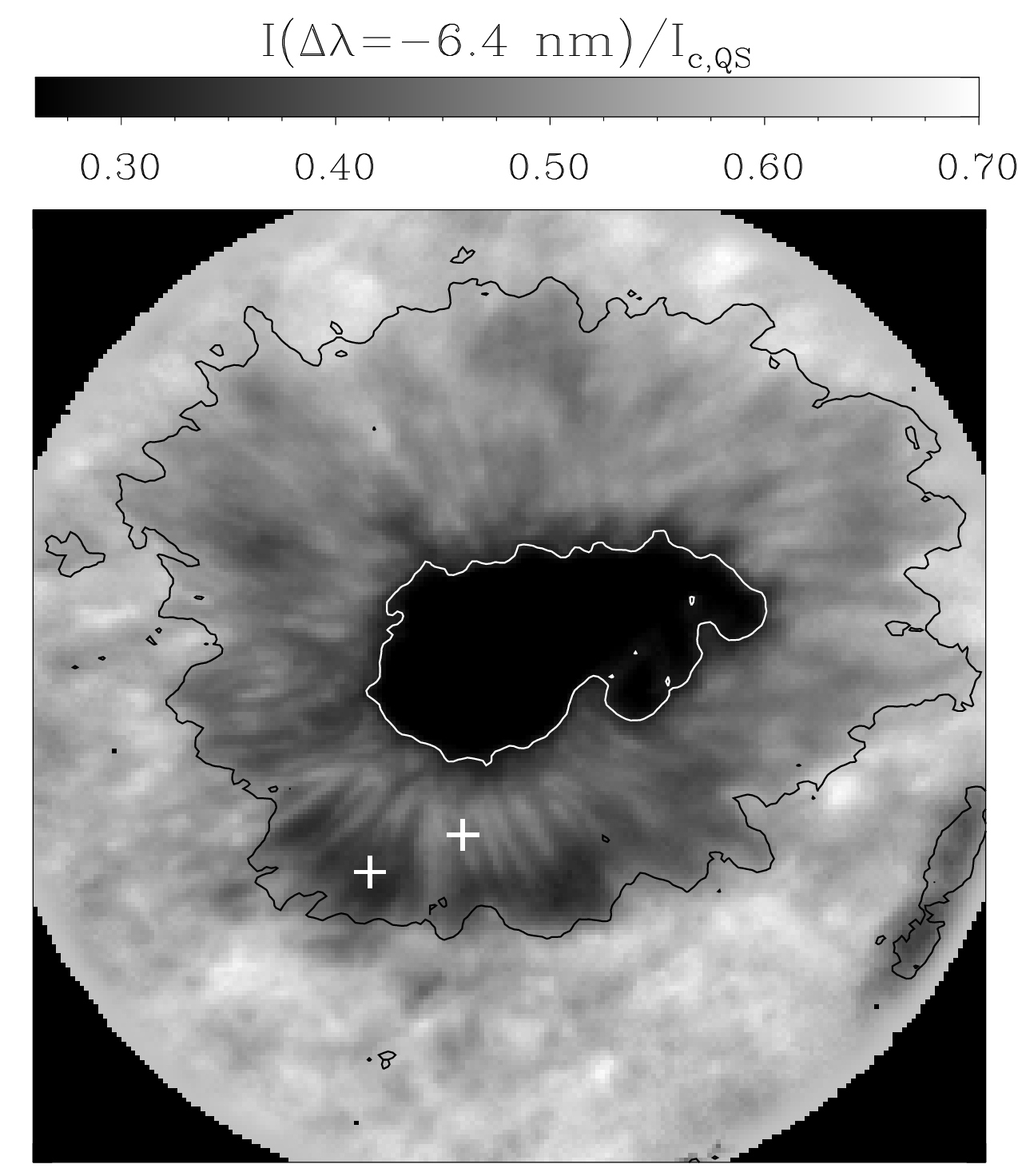}}  
\resizebox{0.49\hsize}{!}{\includegraphics{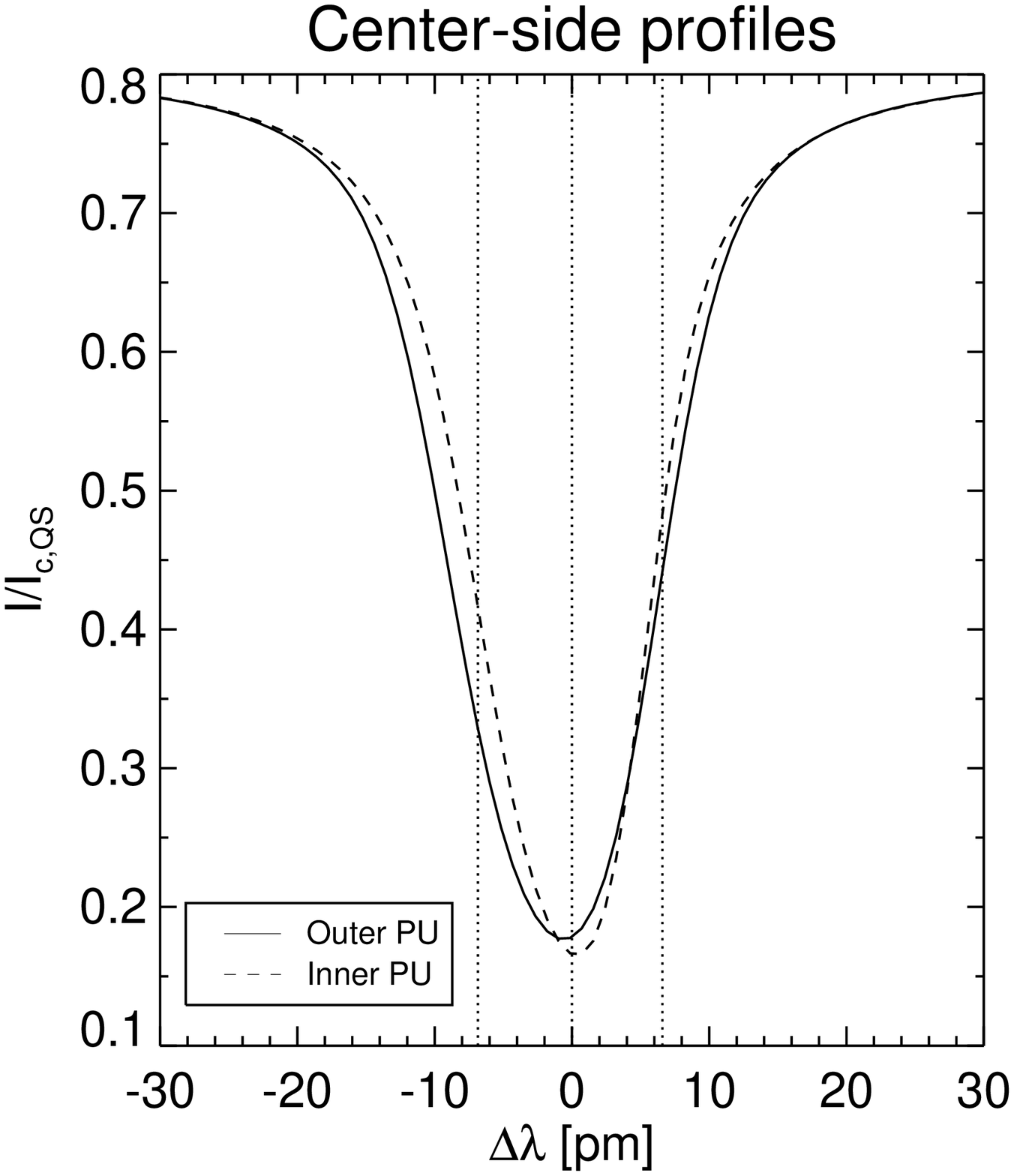}} 
\end{center} 
\caption{ {\em Left:} Intensity map of the spot at $\Delta \lambda = -6.4$ pm
from line center. Two pixels belonging to the inner and outer center-side
penumbra are marked with crosses. {\em Right:} Best-fit intensity profiles
returned by the inversion at the spatial positions marked with crosses in the
left panel. Solid: outer penumbra. Dashed: inner penumbra. The
vertical dotted lines mark fixed wavelength positions in the blue wing, the
line core, and the red wing (from left to right). The blue and red wing
positions correspond to filtergrams \#42 and \#58 of Fig.~3 in Paper~I.
\label{brightring} 
}
\end{figure}

Here we argue that the bright ring is actually produced by (a) the small
Doppler shifts and line asymmetries observed in the inner penumbra, together
with (b) the increase of the line width toward the outer sunspot boundary. The
left panel of Fig.~\ref{brightring} shows an intensity map of the spot at
$\Delta \lambda = -6.4$ pm from line center (filtergram \#42 in Fig.~3 of
Paper~I). The enhanced brightness of the inner center-side penumbra is
apparent in this image. The intensity profiles emerging from the two spatial
positions marked with crosses are displayed in the right panel of
Fig.~\ref{brightring}. As can be seen, these profiles possess very different
properties: in the outer penumbra (solid line) they are broader and have a
stronger asymmetry than in the inner penumbra (dashed line). The bright ring
results from a comparison of the intensities of pixels belonging to the inner
and outer penumbra at fixed wavelengths (dotted vertical lines). Consider, for
example, wavelengths in the blue wing (leftmost vertical line). At this
wavelength, the inner penumbra is clearly brighter than the outer penumbra. In
the red wing (rightmost vertical line), the inner penumbra would still be
brighter, but with a much reduced contrast. This is consistent with the
observations described in Paper~I. Similar arguments apply to profiles
emerging from the limb-side penumbra: because of the smaller line widths and
asymmetries in the inner penumbra, the intensity at fixed wavelengths is
enhanced relative to the outer penumbra, with higher contrasts redward of the
line core.

\begin{figure}
\begin{center}
\resizebox{0.49\hsize}{!}{\includegraphics{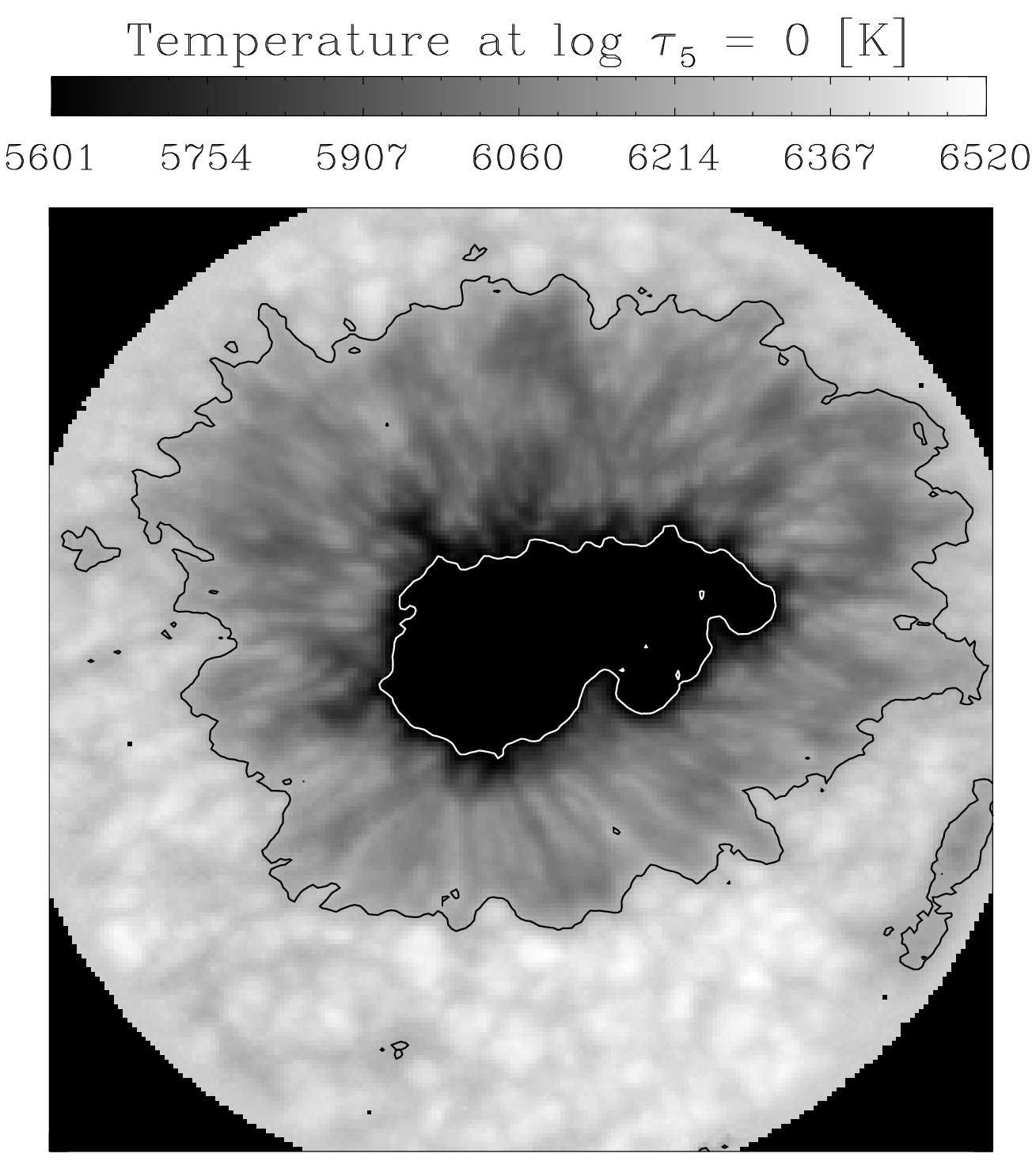}}  
\resizebox{0.49\hsize}{!}{\includegraphics{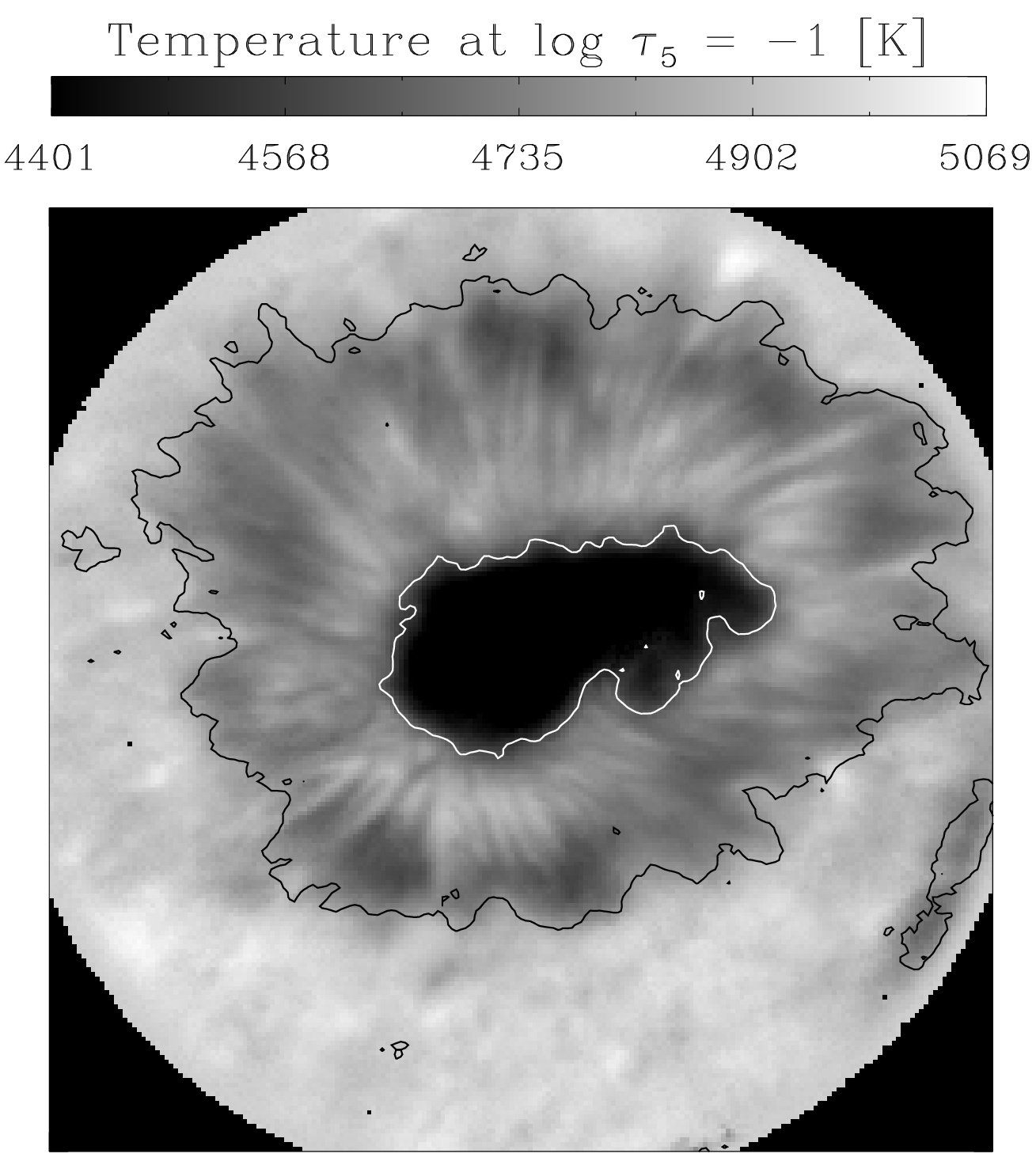}} 
\vspace*{-.5em}
\end{center} 
\caption{Maps of temperature at $\log \tau_{500}= 0$ {\em (left)} and $\log
\tau_{500}= -1$ {\em (right)} resulting from an inversion of the observed
profiles in which only constant velocities along the LOS are permitted.
\label{maps_vel1nodo} 
}
\end{figure}

Thus, the bright ring can be explained satisfactorily in terms of the radial
variation of the microturbulence and the Evershed flow. Microturbulence is an
important parameter: if it were the same everywhere across the spot, redward
of the line core the inner penumbra would appear {\em darker} than the outer
penumbra on the center side, contradicting the observations. To further
support the idea that the bright ring is not due to temperature enhancements,
we have carried out the same inversions by forcing the LOS velocity to be
constant with height (i.e., we do not allow for line asymmetries). In
addition, we set the microturbulence to zero. The thermal stratification
resulting from these inversions is shown in Fig.~\ref{maps_vel1nodo} at $\log
\tau_{500} = 0$ and $-1$. While the temperature in the deep layers is
essentially unchanged, the map of temperature at $\log \tau_{500} = -1$ shows
a nice bright ring in the inner penumbra. From this we conclude that the only
way to explain the enhanced line-wing intensities of the inner penumbra when
no line asymmetries are considered is by an increase of the temperature in the
mid photosphere.

Using flowless maps, Balasubramaniam (2002) detected bright structures in the
inner penumbra at wavelengths close to line center (his Fig.~8). Transforming
observed intensities into radiation temperatures, he concluded that these
structures are hotter than their surroundings. In our opinion, however, they
more likely trace regions of reduced Evershed flows and small-scale
turbulence.

\subsection{Interpretation of the LOS velocity stratifications}
\label{signreversal}

In Paper~II we demonstrated that the uncombed model proposed by Solanki \&
Montavon (1993) is able to explain the bisector shapes observed in the
penumbra of NOAA 10019. In an uncombed penumbra, the Evershed flow occurs
along flux tubes whose magnetic fields are more horizontal than that of their
surroundings. Here we use this model to interpret the LOS velocity
stratifications displayed in Fig.~\ref{averageLOS1}.

As pointed out in Sect.\ \ref{inversion}, it is the magnitude and slope of the
inferred velocity stratification which inform us about the flow geometry (flow
speed and flow angle) and height of the Evershed channels. Figure~5 of
Paper~II shows that, in the very outer penumbra, two flow channels stacked
along the LOS and with different inclination angles are necessary to
understand the kinks and reversals of the observed bisectors. In the inner and
mid penumbra the bisectors are rather linear, and can be explained in terms of
a single, deep-lying flow channel.

\begin{figure}
\begin{center}
\resizebox{0.47\hsize}{!}{\includegraphics{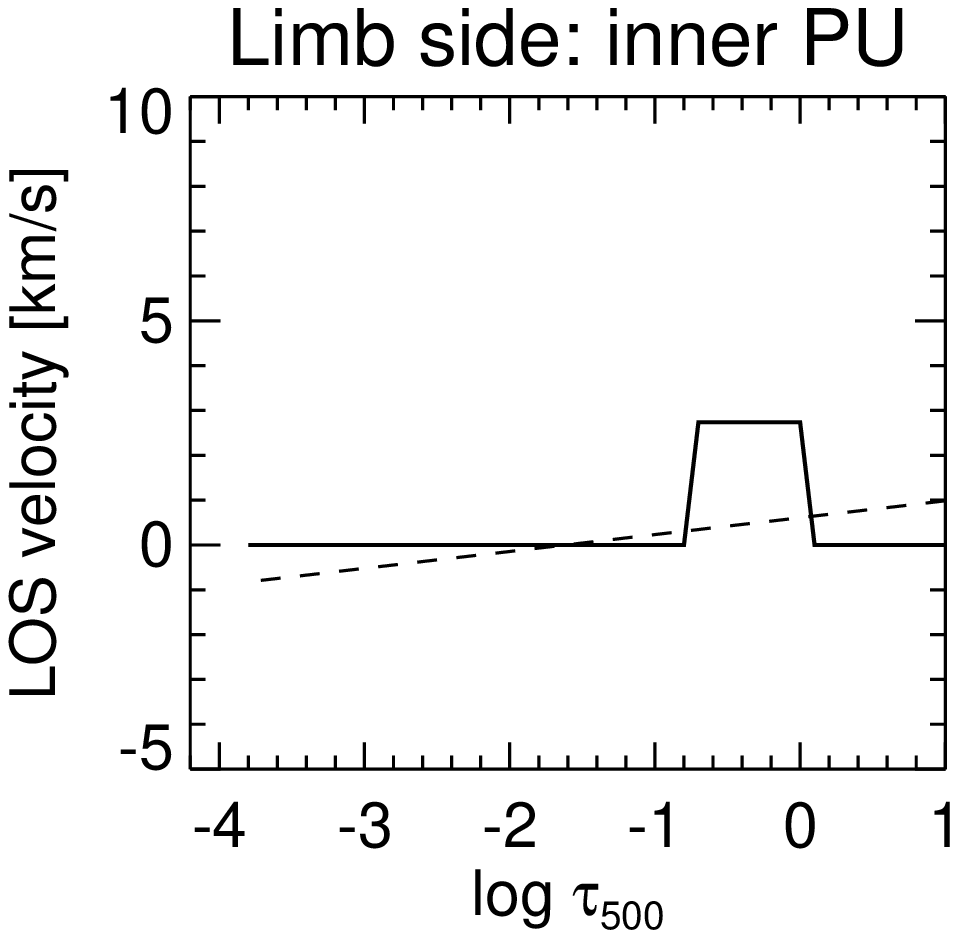}}
\resizebox{0.47\hsize}{!}{\includegraphics{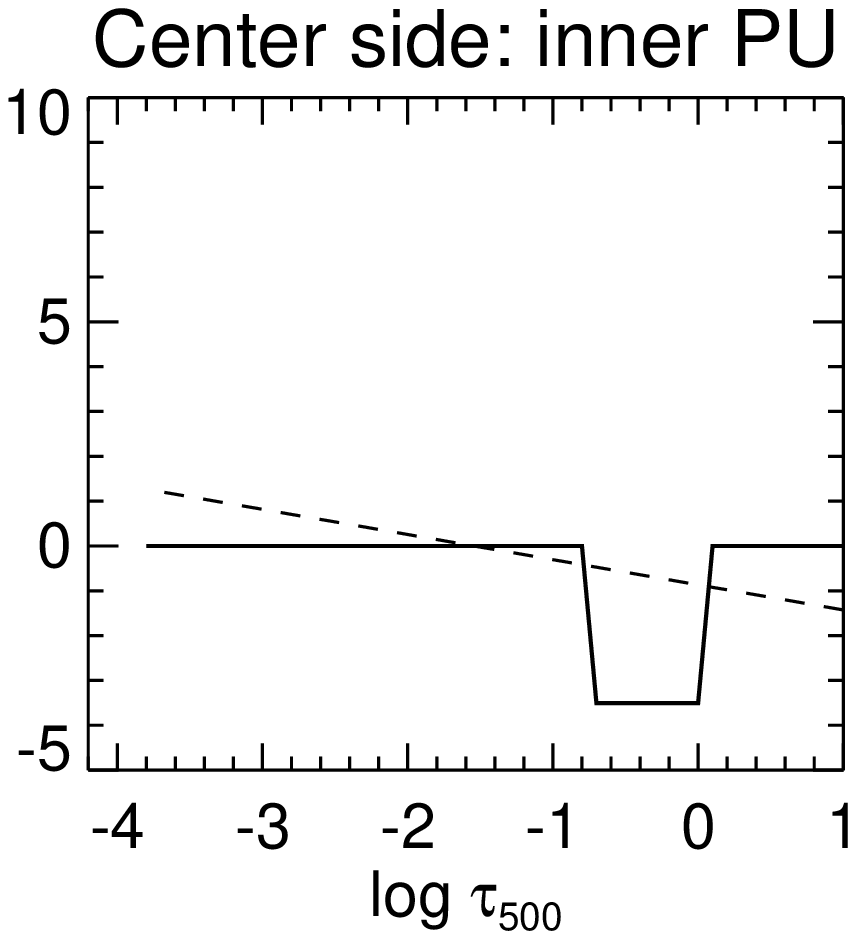}} \\
\resizebox{0.47\hsize}{!}{\includegraphics{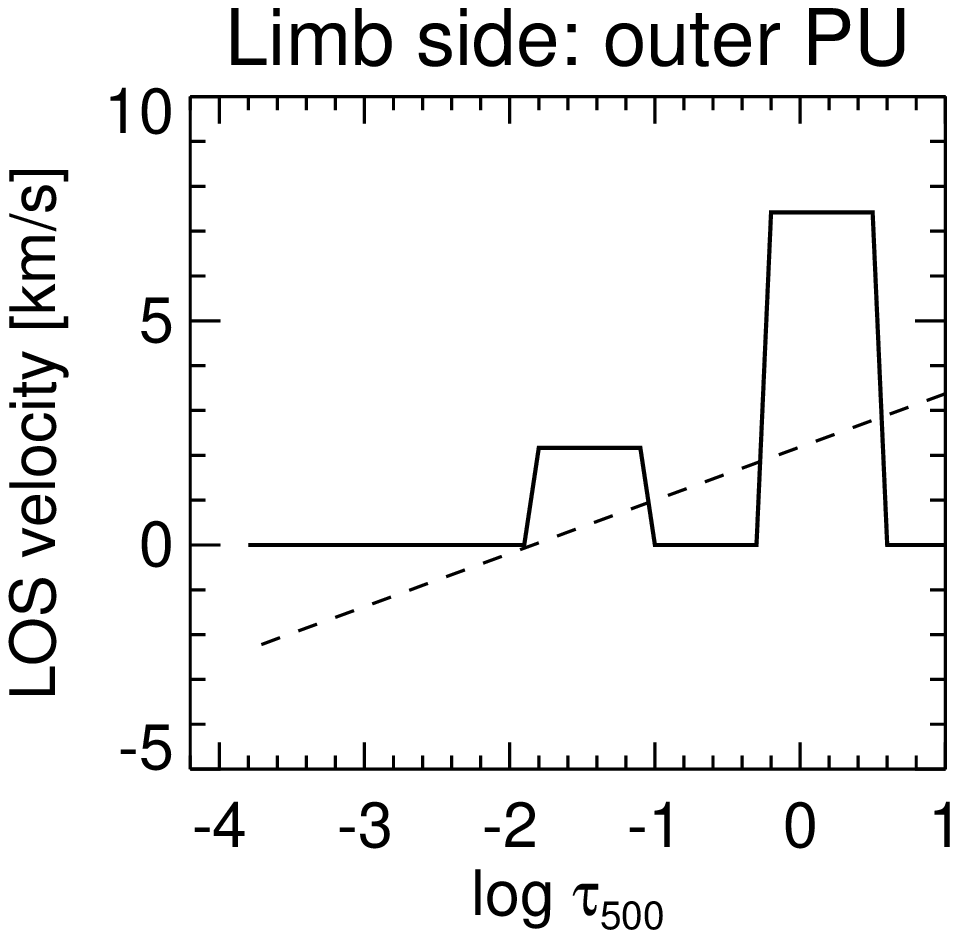}} 
\resizebox{0.47\hsize}{!}{\includegraphics{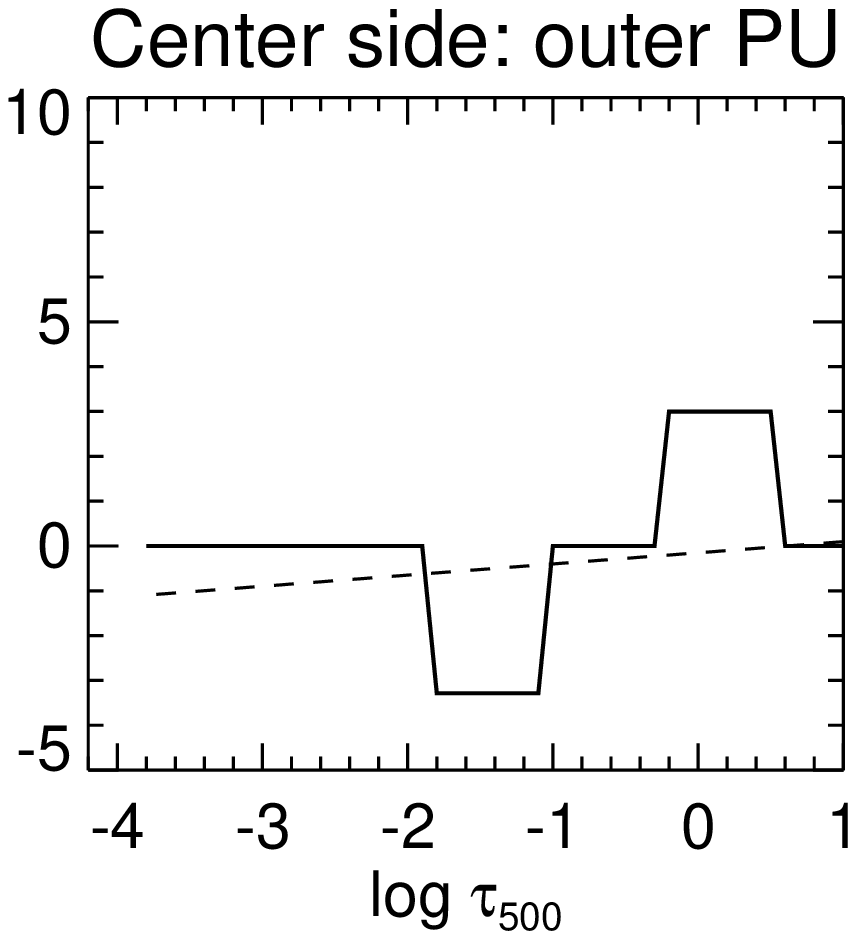}}
\end{center}
\caption{LOS velocity stratifications due
to flux tubes along the LOS (solid lines) that would result in {\em linear} 
velocity stratifications (dashed lines) when two nodes are used to 
retrieve the actual velocity stratification. See text for details on the
flux-tube geometries adopted in the inner and outer penumbra.
\label{averageLOS1_bis}}
\end{figure}

Based on these results, in the inner penumbra ($r=0.6R$) we consider a flow
channel located in deep layers, between $\log \tau_{500}= -0.7$ and 0.  The
flow has a speed of 8 km~s$^{-1}$ and an inclination of $87^\circ$ with
respect to the vertical. In the outer penumbra ($r=0.97R$), we assume two flow
channels located in deep and mid photospheric layers. The deep-lying channel
extends from $\log \tau_{500}= -0.2$ to 0.5. It has a flow speed of 8
km~s$^{-1}$ and an inclination of 135$^\circ$. The upper channel extends
from $\log \tau_{500}= -1.8$ to $-1.1$ and is assumed to have
a flow speed of 7~km~s$^{-1}$ and an inclination of 85$^\circ$. This
flow configuration is similar to the one deduced in Paper II, except for the
fact that our higher-lying channel is slightly inclined upward instead of
being horizontal. The LOS velocities resulting from the two
flux-tube geometries at the heliocentric angle of our observations ($\theta =
23^\circ$) are represented by solid lines in Fig.~\ref{averageLOS1_bis}, for
the center and limb-side penumbra.

To investigate whether or not these flow configurations can explain the LOS
velocities retrieved from the inversion, we have computed synthetic profiles
under the assumption that the resolution element contains two different
atmospheres. The temperature and pressure stratifications of both atmospheres
are taken to be the mean stratifications deduced from the inversion of NOAA
10019. The LOS velocity is set to zero in the first component, whereas for the
second component we use the stratifications of Fig.~\ref{averageLOS1_bis}. In
the inner penumbra we adopt a filling factor of 0.2, i.e., only 20\% of the
resolution element is occupied by flows. In the outer penumbra the filling
factor is 0.5, as in Paper II.  The resulting profiles are
convolved with the instrumental profile of TESOS and a macroturbulent velocity
of 2~km~s$^{-1}$ (inner penumbra) or 1~km~s$^{-1}$ (outer penumbra). Finally,
they are inverted under the same conditions as the real data (two nodes
for velocity, temperature and microturbulence, and a height-independent
macroturbulence). Not unexpectedly, the simulated profiles show the same
bisector shapes and bisector variations across the penumbra as the real
observations.

\begin{figure*}
\begin{center}
\resizebox{.68\hsize}{!}{\includegraphics{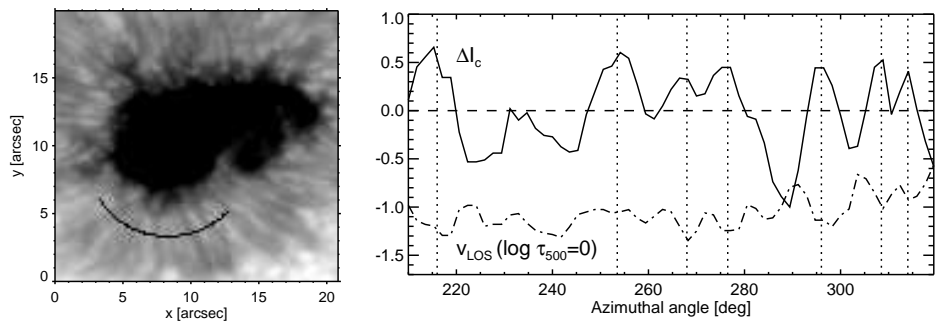}}
\end{center}
\caption{Continuum intensity fluctuations {\em (solid)} and LOS velocity at 
$\log \tau_{500} = 0$ {\em (dashed)} along an azimuthal path crossing the inner 
center-side penumbra as indicated on the left image. Negative velocities
represent blueshifts. The vertical dotted lines mark positions of local 
maxima of the continuum intensity.
\label{corrI_v}
}
\end{figure*}

The {\em linear} velocity stratifications retrieved from the inversion of the
simulated profiles are represented by dashed lines in
Fig.~\ref{averageLOS1_bis}.  The first thing to notice is that the inferred
stratifications change sign in the mid photosphere (around $\log \tau_{500} =
-2$) both on the center and limb sides, even though there are no structures in
high layers that could be responsible for such velocity reversals. Actually,
the sign reversal is the result of extrapolating the jump(s) of velocity
sensed by the LOS to the upper layers, where the response of the line to
plasma motions is small. This extrapolation effect was first described by
Mart\'{\i}nez Pillet (2000) in connection with inversions of Stokes profiles
of visible lines that do not explicitly account for the existence of penumbral
flux tubes. The same effect has been observed by Mathew et al.\ (2003) in
their one-component inversions of near-infrared lines.

The second thing to note is that the azimuthally averaged LOS velocities
inferred from the inversion of NOAA 10019 (Fig.~\ref{averageLOS1}) are in
excellent accord with the {\em linear} velocity stratifications resulting 
from the flux tube configurations displayed in Fig.~\ref{averageLOS1_bis}. 
Specific features that are well reproduced include the following:
\begin{itemize}
\item On the limb side, the slope of the stratifications is larger in the
outer penumbra as compared with the inner penumbra, whereas on the center 
side it is slightly larger in the inner penumbra. 
\item The slope of the stratifications in the inner penumbra is larger on 
the center side.
\item The LOS velocities corresponding to the outer center-side penumbra 
are negative all along the atmosphere.
\end{itemize} 
From this we conclude that the presence of flux tubes in the penumbra can
explain the observed LOS velocity stratifications. In particular, the
seemingly weird run with depth of the LOS velocity in the center-side penumbra
at normalized radial distances larger than 0.9 (with zero or small redshifts
at $\log \tau_{500} = 0$ and blueshifts increasing with height, see
Fig.~\ref{averageLOS1}) can be understood in terms of two flux tubes stacked
along the LOS. The tube lying deeper contributes with positive velocity (i.e.,
a {\em redshift}). This redshift is necessary to explain the bisector
reversals described in Paper II. In order to produce a redshift in the
center-side penumbra, the deep-lying flux tube must be inclined downward with
respect to the horizontal by more than 23$^{\rm o}$ (the heliocentric angle of
the spot), otherwise it would show negative velocities and the slope of the
resulting LOS velocity stratification would be opposite to what we
observe. Flux tubes returning to the solar surface have previously been
inferred by, among others, Westendorp Plaza et al.\ (1997, 2001),
Schlichenmaier \& Schmidt (2000), Bellot Rubio et al.\ (2003, 2004) and
Borrero et al.\ (2004, 2005).

The tube located higher in the atmosphere may be horizontal (as in Paper II)
or slightly inclined upwards (less than 23$^\circ$ from the horizontal). The
later configuration would still produce redshifts on the limb side and
blueshifts on the center side, exactly what is required to explain the
velocity stratifications observed in the penumbra. Interestingly, such
high-lying tubes do not seem to be necessary in the inner penumbra, only in
the mid and outer penumbra.  It is tempting to associate them with those
described by Westendorp Plaza et al.\ (2001) and Bellot Rubio et al.\ (2004).
These authors found indications of a new family of penumbral tubes in the mid
penumbra and beyond. Spectropolarimetric measurements suggest that such tubes
are more vertically oriented than the deeper ones, carrying a fraction of the
Evershed flow to the upper photosphere.

\subsection{Do we see individual flow channels?}
Continuum intensity fluctuations and LOS velocities are very well correlated
in the inner penumbra. This is shown in Fig.~\ref{corrI_v}, where we plot
normalized continuum intensity variations and LOS velocities at $\log
\tau_{500} = 0$ along an azimuthal path crossing 8 distinct continuum
filaments of the inner center-side penumbra (we do not consider the limb-side
part of the azimuthal path because no clear filaments are observed on the limb
side). The curves depicted in Fig.~\ref{corrI_v} reveal that {\em larger
continuum intensities are associated with larger LOS velocities}, i.e., the
Evershed flow is more intense in the brighter structures, at least in the
inner center-side penumbra. This agrees with the conclusions of Bello
Gonz\'alez et al.\ (2005) and Schlichenmaier et al.\ (2005).

However, the most important fact demonstrated by Fig.~\ref{corrI_v} is that
also the dark structures show large LOS velocities. Stray light from the
bright filaments may change the line profile in the dark structures, but only
to some extent (cf.\ Sect. 3.2 in Paper I).  Unless the stray light
contamination is unreasonably large, the velocities found in the dark
filaments must be real. This means that we see the Evershed flow everywhere:
even at a resolution of 0\farcs5, there is almost no position in the penumbra
where the velocity drops to zero (except where it must be zero for geometrical
reasons). In our opinion, this well established fact (see also
Hirzberger \& Kneer 2001 and Rouppe van der Voort 2002) has not been
emphasized adequately in the past.

Since Evershed flows are ubiquitous, it is
tempting to think that the bright and dark structures observed in continuum
images are individual flow channels (i.e., isolated flux tubes).  But this
interpretation is problematic. For example, one would have to explain why the
flow channels possess different temperatures and velocities at the same radial
distance and, more importantly, why there exists such a regular azimuthal
pattern of temperature enhancements/coolings and larger/smaller plasma
flows. An alternative explanation is that the filaments seen in continuum
images are not individual flow channels, but a collection of them. If the
filling factor of the tubes changes regularly in the azimuthal direction,
i.e., if the number of flow channels present in the resolution element changes
azimuthally, then one would observe larger LOS velocities and higher
temperatures at those positions where there are more tubes (provided, of
course, that the tubes are hot in the inner penumbra, as suggested by the
simulations of Schlichenmaier et al.\ 1998 and the uncombed inversions of
Borrero et al.\ 2005). The important idea here is that the intrinsic flow
speed and temperature of the flux tubes would not change azimuthally at a
fixed radial distance from sunspot's center, only the filling factor would
change. We believe this is a natural explanation for the correlation displayed
in Fig.~\ref{corrI_v}.  Independent support for this idea has been gathered
from the analysis of full Stokes profiles of near-infrared lines (Bellot Rubio
2004; Bellot Rubio et al.\ 2004). 

Our results strongly suggest that the filaments observed in Doppler 
maps at a resolution of about 0\farcs5 are still unresolved. That is, 
they are not individual flow channels. For this reason, in Paper~II we 
referred to them as {\em flow filaments}, to clearly distinguish them 
from the actual, smaller flow channels.

\section{Summary}
\label{summary}
Using high spectral and spatial resolution intensity profiles of the
non-magnetic \ion{Fe}{i} 557.6 nm line, we have studied the thermal and
kinematic properties of a sunspot at $23^\circ$ from disk center.  The
profiles, obtained with TESOS, have been inverted in terms of a 
one-component model atmosphere with gradients of the physical quantities. 

The temperature, LOS velocity, and microturbulent velocity maps resulting
from the inversion show that the fine structure of the penumbra is less
apparent in high layers than in deep layers. This suggests that 
the physical mechanism(s) producing the penumbral filaments operate mainly
in the lower photosphere. 

Our analysis confirms the existence of a thermal asymmetry between the center 
and limb side penumbra, the former being hotter by 100-150 K on average. Such 
an asymmetry seems to be related to the presence of a larger number of hot
structures in the outer center-side penumbra, but its nature is still unclear.

We have also investigated the origin of the enhanced line-wing intensities
exhibited by sunspots in the inner penumbra. Our results indicate that the
bright ring is produced by the smaller line widths and Doppler shifts observed
in the inner penumbra as compared with the outer penumbra. Thus, it does not
reflect temperature enhancements in the mid photosphere.

The inferred LOS velocities have been used to determine
the geometry of the Evershed flow assuming axial symmetry of the flow
field. We find that both the flow speed and flow angle increase with optical
depth and radial distance.  Downflows are detected in the mid and outer
penumbra, but only in deep layers ($\log \tau_{500} \leq -1.4$).  By means of
simulated profiles, we have shown that the velocity stratifications retrieved
from the inversion are consistent with the existence of penumbral flux tubes
channeling the Evershed flow. In the inner penumbra, a deep-lying tube
slightly inclined upwards is able to explain the observations. In the outer
penumbra, two flux tubes stacked along the LOS and with different inclination
angles are necessary. The deep-lying tube is inclined downwards, whereas the
higher-lying tube carries upflows. We suggest that the later may belong to a
different family of tubes, perhaps the ones first described by Westendorp
Plaza et al.\ (2001). 

Finally, we have demonstrated that there is an excellent correlation between
continuum intensities and LOS velocities in the inner center-side penumbra,
with larger Evershed flows associated with brighter continuum structures.
Very remarkably, however, dark structures also exhibit significant Evershed
flows. This led us to propose that the bright and dark filaments seen in
continuum images at a resolution of 0\farcs5 are not individual flow 
channels, but a collection of them. 

Our analysis shows that spectroscopy with 2D filter instruments provides 
a wealth of information on the physical properties of sunspots.  Obviously, 
the next step is to carry out 2D full spectropolarimetric observations near the
diffraction limit. This will soon be possible with instruments like the
Interferometric BIdimensional Spectrometer (IBIS) and the KIS/IAA Visible
Imaging Polarimeter (VIP) attached to TESOS.

\begin{acknowledgements}
We thank D.\ Soltau, T.\ Berkefeld and T.\ Schelenz for developing the
Kiepenheuer Adaptive Optics System (KAOS). T.\ Berkefeld operated KAOS during our
observations. T.\ Kentischer built, upgraded, programmed, and aligned TESOS.
This work has been supported by the Spanish MCyT under {\em Programa Ram\'on y
Cajal} and project ESP2003-07735-C04-03, and by the German DFG under grants
SCHL 514/2--1 and PE 782/4. The German Vacuum Tower Telescope (VTT) is
operated by the Kiepenheuer-Institut f\"ur Sonnenphysik on the Spanish
Observatorio del Teide of the Instituto de Astrof\'{\i}sica de Canarias.
\end{acknowledgements}


\begin{thebibliography}{}


\bibitem{} Balasubramaniam, K.S.\ 2002, \apj, 575, 553

\bibitem{} Barklem, P.S., Piskunov, N., \& O'Mara, B.J.\ 2000, \aaps, 142, 467

\bibitem{} Bello Gonz\'alez, N., Okunev, O.V., Dom\'{\i}nguez Cerde\~na, I., 
Kneer, F., \& Puschmann, K.G.\ 2005, \aap, 434, 317

\bibitem{}
  Bellot Rubio, L.R.\ 2003, ASP Conf.\ Ser., 307, 301

\bibitem{}
  Bellot Rubio, L.R.\ 2004, Reviews in Modern Astronomy, 17, 21

\bibitem[Bellot Rubio, Balthasar, Collados, \& 
Schlichenmaier(2003)]{2003A&A...403L..47B} Bellot Rubio, L.R., Balthasar, 
H., Collados, M., \& Schli\-chen\-maier, R.\ 2003, \aap, 403, L47 

\bibitem{} Bellot Rubio, L.R., Balthasar, H., \& Collados, M.\ 2004, \aap,
427, 319

\bibitem{} Borrero, J.M., Solanki, S.K., Bellot Rubio, L.R., Lagg, A., \&
Mathew, S.K.\ 2004, \aap, 422, 1093

\bibitem[Borrero et al.(2005)]{2005A&A...436..333B} Borrero, J.M., Lagg, A.,
Solanki, S.K., \& Collados, M.\ 2005, \aap, 436, 333
 
\bibitem{} Cabrera Solana, D., Bellot Rubio, L.R., \& del Toro Iniesta, 
J.C.\ 2005, \aap, 439, 687

\bibitem{} Georgakilas, A.A., Christopoulou, E.B., Skodras, A., \& Koutchmy,
S.\ 2003, \aap, 403, 1123

\bibitem{} Gray, D.F., 1988, {\em The observation and analysis of stellar
photospheres}, (Wiley: New York)

\bibitem{} Hirzberger, J., \& Kneer, F.\ 2001, \aap, 378, 2001

\bibitem{} Kentischer, T.J., Schmidt, W., Sigwarth, M., \& Uexkuell, M.V.\
1998, \aap, 340, 569


\bibitem{} Maltby, P.\ 1964, Astrophysica Norvegica, 8, 205

\bibitem[Mart{\'{\i}}nez Pillet(2000)]{2000A&A...361..734M} Mart\'{\i}nez 
Pillet, V.\ 2000, \aap, 361, 734 


\bibitem{} Mathew, S.K., Lagg, A., Solanki, S.K., Collados, M., Borrero, J.M.,
et al.\ 2003, \aap, 410, 695

\bibitem[Mathew et al.(2004)]{2004A&A...422..693M} Mathew, S.K., Solanki, 
S.K., Lagg, A., Collados, M., Borrero, J.M., et al.\ 2004,  \aap, 422, 693 
 
\bibitem{} Rimmele, T.R.\ 1995, \aap, 298, 260

\bibitem{} Rouppe van der Voort, L.H.M.\ 2002, A\&A, 389, 1020 

\bibitem[Rouppe van der Voort, L{\" o}fdahl, Kiselman, \& 
Scharmer(2004)]{2004A&A...414..717R} Rouppe van der Voort, L.H.M., 
L\"ofdahl, M.G., Kiselman, D., \& Scharmer, G.B.\ 2004, \aap, 414, 717 

\bibitem{} Ruiz Cobo, B., \& del Toro Iniesta, J.C.\ 1992, \apj, 398, 375

\bibitem{} Ruiz Cobo, B., \& del Toro Iniesta, J.C.\ 1994, \aap, 283, 129

\bibitem{} S\'anchez Almeida, J., Ruiz Cobo, B., \& del Toro Iniesta, J.C.,
  1998, \aap, 314, 295

\bibitem{} S\'anchez Cuberes, M., Puschmann, K.G., \& Wiehr, E.\ 2005, \aap, 
440, 345

\bibitem[Scharmer et al.(2002)]{2002Natur.420..151S} Scharmer, G.B., 
Gudiksen, B.V., Kiselman, D., L{\"o}fdahl, M.G., \& Rouppe van der Voort, 
L.H.M.\ 2002, \nat, 420, 151 
 
\bibitem{} Schlichenmaier, R., \& Collados, M.\ 2002, \aap, 381, 668

\bibitem{}  Schlichenmaier, R., \& Schmidt, W.\ 2000, \aap, 358, 1122

\bibitem{} Schlichenmaier, R., Bellot Rubio, L.R., \& Tritschler, A.\ 2004,
\aap, 415, (Paper II)

\bibitem{} Schlichenmaier, R., Bellot Rubio, L.R., \& Tritschler, A.\ 2005,
Astron.\ Nachr., 326, 301

\bibitem[Schlichenmaier, Bruls, \& Schussler(1999)]{1999A&A...349..961S} 
Schlichenmaier, R., Bruls, J.H.M.J., \& Sch\"ussler, M.\ 1999, \aap, 349, 961 

\bibitem{} Schlichenmaier, R., Jahn, K., \& Schmidt, H.U.\ 1998, \aap, 337, 897


\bibitem[Schmidt \& Fritz(2004)]{2004A&A...421..735S} Schmidt, W., \& 
Fritz, G.\ 2004, \aap, 421, 735 
 
\bibitem{} Solanki, S.K.\ 2003, \aapr, 11, 153

\bibitem[Solanki \& Montavon(1993)]{1993A&A...275..283S} Solanki, S.K., \& 
Montavon, C.A.P.\ 1993, \aap, 275, 283 
 
\bibitem{} Soltau, D., Berkefeld, T., von der L\"uhe, O., W\"oger, F., \&
Schelenz, T.\ 2002, Astron.\ Nachr., 323, 236

\bibitem{} Thomas, J.H., \& Weiss, N.O.\ 2004, \araa, 42, 517

\bibitem{} del Toro Iniesta, J.C., 2003, {\em Introduction to Spectropolarimetry}, 
(Cambridge University Press: Cambridge)

\bibitem[ttr]{} del Toro Iniesta, J.C., Tarbell, T.D., \& Ruiz Cobo, B.\ 1994, 
\apj, 436, 400 (TTR)

\bibitem[Tritschler et al.(2004)]{2004A&A...415..717T} Tritschler, A.,
Schlichenmaier, R., Bellot Rubio, L.R., and the KAOS Team 2004, \aap, 415, 717
(Paper I)

\bibitem{} Tritschler, A., Schmidt, W., Langhans, K., \& Kentischer, T.J.\
2002, Solar Phys., 211, 17

\bibitem{} von der L\"uhe, O., Soltau, D., Berkefeld, T., \& Schelenz, T.\
2003, SPIE, 4853, 187

\bibitem[Westendorp Plaza et al.(1997)]{1997Natur.389...47W} Westendorp 
Plaza, C., del Toro Iniesta, J.C., Ruiz Cobo, B., Mart\'{\i}nez Pillet, V., 
Lites, B.W., \& Skumanich, A.\ 1997, \nat, 389, 47 
 
\bibitem{} Westendorp Plaza, C., del Toro Iniesta, J.C., Ruiz Cobo, B.,
Mart\'{\i}nez Pillet, V., Lites, B.W., \& Skumanich, A.\ 2001, \apj, 547,
1130

\end{thebibliography}
\end{document}